\documentclass[11pt,reqno]{amsart}
\pdfoutput=1 
\usepackage[a4paper, portrait, margin=1in]{geometry}
\usepackage[english]{babel}
\usepackage[utf8]{inputenc}
\usepackage[T1]{fontenc}
\usepackage{helvet}
\usepackage{etoolbox}
\usepackage{graphicx}
\usepackage{caption}
\usepackage{booktabs}
\usepackage{xcolor} 
\usepackage[colorlinks, citecolor=cyan]{hyperref}
\usepackage{caption}
\graphicspath{ {./images/} }
\usepackage{scrextend}
\usepackage{fancyhdr}
\usepackage{graphicx}
\usepackage{amsthm}
\usepackage{tikz}
\usepackage{caption}
\usepackage{subcaption}

\usepackage{mathtools}
\usepackage{amsmath,amsfonts,amssymb,amsthm,epsfig,epstopdf,url,array,dsfont}
\usepackage{thmtools}
\usepackage[capitalise]{cleveref}
\usepackage{graphicx}

\usepackage{amsmath} 
\usepackage{amssymb} 
\usepackage[english]{babel}
\usepackage{subcaption}
\usepackage{float}
\usepackage{here}
\usepackage[all]{xy}
\usepackage[scr=rsfs,cal=boondox]{mathalfa}

\newtheorem{theorem}{Theorem}[section]
\newtheorem{lemma}[theorem]{Lemma}

\newtheorem{corollary}[theorem]{Corollary}

\newtheorem{proposition}[theorem]{Proposition}
\newtheorem{definition}[theorem]{Definition}

\theoremstyle{definition}\newtheorem{rem}[theorem]{Remark}

\theoremstyle{definition}\newtheorem{example}[theorem]{Example}

\usepackage{comment}
\usepackage{bbold}
\usepackage{pgfplots}
\pgfplotsset{compat=1.11}
\usepgfplotslibrary{fillbetween}
\usetikzlibrary{intersections}
\pgfdeclarelayer{bg}
\pgfsetlayers{bg,main}

\usepackage{amsmath,amssymb,xparse} % xparse gives us nice argument-parsing tools

\RenewDocumentCommand{\L}{>{\SplitArgument{1}{,}}m}
  {%
    \Laux #1%
  }

\NewDocumentCommand{\Laux}{mm}
  {%
    \operatorname{L}\!\bigl(
      \mathbb{C}^{#1}%
      \IfNoValueTF{#2}
        {}%                       <-- nothing if absent
        {\otimes\mathbb{C}^{#2}}% <-- tensor factor if present
    \bigr)%
  }

\NewDocumentCommand{\U}{>{\SplitArgument{1}{,}}m}
  {%
    \Uaux #1%
  }

\NewDocumentCommand{\Uaux}{mm}
  {%
    \operatorname{U}\!\bigl(
      \mathbb{C}^{#1}%
      \IfNoValueTF{#2}
        {}%                       <-- nothing if absent
        {\otimes\mathbb{C}^{#2}}% <-- tensor factor if present
    \bigr)%
  }

\newcommand{\trsum}[1]{\operatorname{tr}_1#1\oplus\operatorname{tr}_2#1}

\newcommand{\lambdatrsum}[1]{\lambda(\operatorname{tr}_1#1\oplus\operatorname{tr}_2#1)}

\newcommand{\RR}{\mathbb{R}}

\newcommand{\lambdasum}[1]{\lambda^{(1, 2)}(#1)}

\newcommand{\unorm}[1]{\left\vert\hspace*{-1.2pt}\left\vert\hspace*{-1.2pt}\left\vert #1 \right\vert\hspace*{-1.2pt}\right\vert\hspace*{-1.2pt}\right\vert}

\newcommand{\lambdaone}{\lambda^{(1)}}
\newcommand{\lambdatwo}{\lambda^{(2)}}

\newcommand{\be}{\begin{equation}}
\newcommand{\ee}{\end{equation}}
\newcommand{\CCC}{\mathbb{C}}

\newcommand{\RRR}{\mathbb{R}}

\DeclareMathOperator{\tr}{tr}
\DeclareMathOperator{\diag}{diag}
\DeclareMathOperator{\id}{id}
\DeclareMathOperator{\rk}{rank}
\DeclareMathOperator{\sign}{sign}

\DeclareMathOperator{\cH}{\mathcal{H}}

\renewcommand{\1}{\mathds{1}}

\makeatletter

\def\section{\@startsection{section}{1}%
  {0pt}%
  {12pt plus 4pt minus 2pt}%
  {1sp plus 2pt minus 2pt}%
  {\normalfont\fontsize{11}{15}\selectfont\bfseries}}

\def\subsection{\@startsection{subsection}{2}%
  {12pt}%
  {12pt plus 4pt minus 2pt}%
  {1sp plus 2pt minus 2pt}%
  {\normalfont\fontsize{11}{15}\selectfont\bfseries}}

\makeatother

\usepackage{hyperref}

\date{}    

\begin{document}

\title{\textbf{ Sharp Inequalities for Schur-Convex Functionals of Partial Traces over Unitary Orbits}\\
	}

\author[Rico]{ Pablo Costa Rico$^{1,2}$}
\email{pablo.costa@tum.de}
\author[Shteyner]{ Pavel Shteyner$^{3,4}$}
\email{p.shteyner@campus.technion.ac.il}
\address{$^1$ Department of Mathematics, Technical University of Munich, Garching bei München, Germany}
\address{$^2$ Munich Center for Quantum
Science and Technology (MCQST),  M\"unchen, Germany}
\address{$^3$ Department of Mathematics, Bar-Ilan University, Ramat Gan, Israel}
\address{$^4$ Department of Mathematics, Technion - Israel Institute of Technology, Israel}

\begin{abstract} 

While many bounds have been proved for partial trace inequalities over the last decades for a large variety of quantities,  recent problems in quantum information theory demand sharper bounds. In this work, we study optimal  bounds for partial trace quantities in terms of the spectrum; equivalently, we determine the best bounds attainable over unitary orbits of matrices. We solve this question for Schur-convex functionals acting on a single partial trace in terms of eigenvalues for self-adjoint matrices and then we extend these results to singular values of general matrices. We subsequently extend the study to Schur-convex functionals that act on several partial traces simultaneously and present sufficient conditions for sharpness. In cases where closed-form maximizers cannot be identified, we present quadratic programs that yield new computable upper bounds for any Schur-convex functional. We additionally present examples demonstrating improvements over previously known bounds. Finally, we conclude with the study of optimal bounds for an $n$-qubit system and its subsystems of dimension $2$.
\end{abstract}

\maketitle

%\begingroup
\setcounter{tocdepth}{1}
\tableofcontents
%\endgroup

\newpage

\section{Introduction}

The partial trace is a fundamental operation in quantum theory that allows one to extract information of a subsystem from a larger composite system. In quantum theory, this information is encoded in terms of  quantum states, represented by positive semidefinite matrices of unit trace acting on a Hilbert space. One of the central topics of study is the relationship between the reduced states of a subsystem and the properties of the global state.

In the particular field of quantum information theory, the partial trace provides the mathematical tool for defining key quantities such as conditional entropies,  (conditional) mutual information or  entanglement measures \cite{lieb1973proof,vedral1998entanglement,wehrl1978general} which allow one to quantify correlations, conditional dependence or entanglement properties among subsystems. Furthermore, it is also a key ingredient in dilation theorems such as Stinespring's theorem \cite{stinespring1955positive} and plays an important role in new characterizations of long-standing open problems, such as the distillability of Werner states \cite{horodecki2022five,rico2025new}. In the last decades, the study of information-theoretic quantities has  yielded a  broad literature of partial trace inequalities at the operator and at the scalar level \cite{Audenaert.2007,choi2018inequalities,eltschka2018exponentially,lieb1973proof,Lin.2016,lin2023new,Rains.2000,Rastegin.2012,rico2025new,rico2025partial}. However, recent problems like the distillability of Werner states \cite{rico2025new} demand  sharper bounds that relate the spectra and singular values of a matrix and  its partial traces.

For the case of quantum states, the relation between the spectrum of a matrix and the spectrum of its partial traces  is not arbitrary. This is the essence of one of the major problems in quantum information theory known as the quantum marginal problem, solved after many years of work \cite{Bravyi.2003,higuchi2003one,higuchi2003one2,klyachko2002coherent,klyachko2004quantum}. In 2004, Klyachko  provided a complete characterization by using techniques of symplectic geometry \cite{klyachko2004quantum}. He showed that the relations between the spectra of a matrix and its partial traces consist of finitely many linear inequalities, turning the geometry of the combined spectra into a convex polytope.

However, performing convex optimization on these polytopes is in general not practical, since the constraints vary with the system size and are highly nontrivial \cite{klyachko2004quantum}. Therefore, in this work we relax the set of constraints and, relying on  majorization theory, we derive necessary and sufficient spectral  constraints ensuring that the spectra of the partial traces are majorized by those of diagonal matrices on the unitary orbit. This allows the application of Schur-convex and Schur-concave functionals providing sharp estimates in terms of the spectra and singular values.

\subsection{Main results}

In this work we are concerned with   optimization problems for Schur-convex functionals of partial traces over unitary orbits. Our focus is on understanding the extremal behavior of
\begin{equation*}
   \tr_i[UCU^*] \quad \text{and} \quad \tr_i[UCV]\, , \quad i=1,2\, , 
\end{equation*}
where $C\in \L{d_1,d_2}$ and the optimizations are taken over unitaries in  $U,V \in \U{d_1,d_2}$. Throughout this paper, we will denote by $f$  a Schur-convex functional.  We begin in Section \ref{sec:DiagonalizationPartialTraces}, where we relate the eigenvalues of the partial traces of a self-adjoint matrix to the diagonal values of a matrix in its unitary orbit. This observation will be key throughout the entire work, and will allow us  to show in Proposition \ref{prop:diagonalization} that majorization for eigenvalues of partial traces is equivalent to the majorization of its diagonals over unitary orbits.

In Section \ref{sec:SinglePartialTrace} we analyze the optimization problems:
\begin{equation*}
    \max_{U \in \U{d_1,d_2}}f(\tr_i[UCU^*])\, , \qquad \text{and} \qquad \max_{U,V \in \U{d_1,d_2}}f(\tr_i[UCV])  \, ,
\end{equation*}
$i=1,2$. When $C$ is self-adjoint, we show in Proposition \ref{MajorizationPartialTrace} that for the second partial trace the maximum over the orbit $U \cdot U^*$ is obtained by arranging the eigenvalues of 
$C$ in the decreasing order in the diagonal basis. To obtain the result for the first partial trace, we consider the flip operator $F:\mathbb{C}^{d_1}\otimes \mathbb{C}^{d_2} \to \mathbb{C}^{d_2}\otimes \mathbb{C}^{d_1}$, which is a linear map,  but, when $d_1\neq d_2$, is not an endomorphism, since it does not preserve the tensor structure.  It further satisfies $F^* F=I_{\CCC^{d_1}\otimes  \CCC^{d_2}}$ and $FF^*  =I_{\CCC^{d_2}\otimes  \CCC^{d_1}}$. Writing $\Lambda$
 for the matrix of the eigenvalues of $C$ in the decreasing order, we prove that\footnote{Note that $F^* \Lambda F$ is still a diagonal matrix from the unitary orbit of $C$.}
\begin{equation*}
    \max_{U \in \U{d_1,d_2}}f(\tr_1[UCU^*])=f(\tr_1[ F^* \Lambda F])\, , \quad \text{and} \quad \max_{U \in \U{d_1,d_2}}f(\tr_2[UCU^*])=f(\tr_2  [\Lambda] )\, .
\end{equation*}
These identities follow from majorization relations for partial traces over unitary orbits.

In addition, for a general $C$, we solve the second optimization problem for monotonically increasing  Schur-convex functionals $f$ in \cref{theo:WeakMajorizationSingularValues}  finding:
\begin{equation*}
    \max_{U,V \in \U{d_1,d_2}}f(\tr_1[UCV])=f(\tr_1[ F^* \Sigma F])\, , \quad \text{and} \quad \max_{U,V \in \U{d_1,d_2}}f(\tr_2[UCV])=f(\tr_2  [\Sigma] )\, ,
\end{equation*}
where $\Sigma$ is the singular value matrix associated to $C$ with singular values ordered in the decreasing order.

In the second part of Section  \ref{sec:SinglePartialTrace}, we will look into the minima for the case of positive semidefinite matrices. When the local dimensions coincide, $d_1=d_2$, the minimum is attained at the same basis: the Bell basis.  Hence, these  bounds (the maximum and the minimum) are sharp when no information about the underlying basis is available. Therefore, when $d_1=d_2$, we provide sharp bounds for maxima and minima of Schur-convex functionals acting on the partial traces of positive semidefinite matrices. This is particularly relevant in quantum information theory, since quantum states are in particular positive semidefinite matrices.

In  Section \ref{sec:sufficient2pt}, we investigate the combined spectra of both partial traces simultaneously, by looking into the optimization problem
\begin{equation*}
     \max_{U \in \U{d_1,d_2}}f\left(\tr_1[UCU^*]\oplus \tr_2[UCU^*]\right)\, ,
\end{equation*}
 Specifically, in the first part of Section \ref{sec:sufficient2pt} we will consider the case $d_1=d_2$, and afterwards, we will tackle the general case. Under some specific assumptions, we prove that the maximum is attained at the diagonal matrix with the eigenvalues in the  decreasing order, see Propositions \ref{prop:NewSufficientConditionsSquareCase}, \ref{prop:sufficient_conditions_small_dimensions}, \ref{prop:SufficientCaseDifferentDimensions},  \ref{prop:sufficientRank3}, and Corollary \ref{coro:SufficientConditionsd<4SquareCase}. However, this is in general not true as we show in Proposition \ref{prop:ImpossibleSquareGeneralRanks}, Corollary \ref{coro:SmallRanksSquareCase} or Lemmas \ref{lem:rankBetweenMultipliesd_2} and \ref{lema:r<=d_2}.

Since the sufficient conditions guaranteeing optimality in a diagonal matrix are quite restrictive, in Section \ref{sec:QP} we construct  quadratic programs (QPs) that incorporate these restrictions and  allow us to obtain new upper bounds for the joint optimization problem for any self-adjoint matrix. Although these bounds are not sharp, they improve upon previously known bounds. In Example \ref{example:numerics}, we show how the solutions of the QP can improve upon previous  well-known results for $p$-norms or for the $\alpha$-Rényi entropy.

In parallel with the eigenvalue analysis, we study the joint optimization problem for the singular values:
\begin{equation*}
     \max_{U,V \in \U{d_1,d_2}}f\left(\tr_1[UCV]\oplus \tr_2[UCV]\right)\, ,
\end{equation*}
for any matrix $C \in \L{d_1,d_2}$. In Section \ref{sec:SVD}, we will provide sufficient conditions for the maximum to be attained at the diagonal matrix with singular values in the decreasing order. Moreover, we will also extend our quadratic programming to cover this scenario, enabling us to obtain bounds for any vector of singular values.

Finally, in Section \ref{sec:n-partite} we extend our analysis to multipartite qubit systems. For a self-adjoint operator $C \in L((\mathbb{C}^2)^{\otimes n})$ with vector of eigenvalues $\lambda \in \mathbb{R}^{2n}$ in the decreasing order satisfying $\lambda_4=\hdots=\lambda_{2^n-3}$, we prove
\begin{equation*}
    \max_{U \in \operatorname{U}((\mathbb{C}^2)^{\otimes n})}f\left(\bigoplus_{j=1}^n\tr_{\{1,\hdots,n\}\setminus \{j\}} [UCU^*] \right)=f\left(\bigoplus_{j=1}^n\tr_{\{1,\hdots,n\}\setminus \{j\}} [\Lambda] \right)\, .
\end{equation*}
This leads to sharp bounds for various entropic information quantities associated with 
$n$-qubit systems.

\section{Notation and Preliminaries}

In this work, we will denote by $\L{d}$ the set of linear operators on $\mathbb{C}^d$ and by $\U{d}$ the set of unitary operators on $\mathbb{C}^d$. Given a self-adjoint matrix $C \in \L{d}$, we will denote by $c_i$ the diagonal entries for $i=1,\hdots, d$ and write $\diag(C)=(c_1,\hdots, c_d)^T\in \mathbb{R}^d$, where $T$ denotes the transpose. Conversely, given any vector $c$ we will denote by $\diag(c)$ the diagonal matrix with  vector $c$ on the main diagonal. This dual use of the $\diag$ notation will be clear from the context.

\subsection{Partial traces}

The partial trace is an operation that traces over one or many subsystems of a composite system. In bipartite systems, this operation can be conceived in two equivalent ways:

On the one hand, given $C \in \L{d_1,d_2}$, we can think of this matrix as a $d_1\times d_1$ block matrix $[C_{ij}]$ with blocks of size $d_2\times d_2$. The partial traces of $C$ are then given by
\begin{equation}\label{eq:computationPartialTrace}
    \tr_1 [C]=\sum_{i=1}^{d_1} [C_{ii}] \in L(\mathbb{C}^{d_2}), \quad (\tr_2 [C])_{ij}=\tr[C_{ij}] \in L(\mathbb{C}^{d_1}).
\end{equation}
As an example, we can write the diagonal entries of partial traces in terms of the diagonal entries of the original matrix $C$,
\begin{equation}\label{eq:diag:one,two}
\diag(\tr_1 [C])_i= \sum_{s=1}^{d_1}c_{(s-1)d_2+i}, \quad i=1,\dots,d_2 \, ,
\quad \diag(\tr_2 [C])_j=\sum_{s=1}^{d_2} c_{(j-1)d_2+s}, \quad j=1,\dots,d_1 \, ,
\end{equation}
These operations can also be extended to vectors $c \in \mathbb{R}^{d_1d_2}$ by writing them as a diagonal of a matrix with off-diagonal zeros. To show clearly the structure, we will often write them as
\begin{equation}
    c^T=\bigoplus_{i=1}^{d_1}(\underbrace{c_{(i-1)d_2+1}, \hdots, c_{id_2}}_{d_2})\, .
\end{equation}

 On the other hand, the partial traces $\tr_1$  and $\tr_2$ can also be considered as completely positive linear maps $\tr_1:\L{d_1, d_2} \to \L{d_2}$, $\tr_2:\L{d_1, d_2} \to \L{d_1}$, which are defined as the unique linear  maps satisfying: 
 \begin{equation}
   \tr[C(\1\otimes T)]=\tr[\tr_1[C]T]   \, , \quad \tr[C(S\otimes \1)]=\tr[\tr_2[C]S] \, , 
 \end{equation}
  for every $T \in \L{d_2}$ and $S \in \L{d_1}$.

 The relation between the first partial trace and the second partial trace is given through the flip (or swap) map, which is the linear  extension of the operator  acting on tensor products as $F(x \otimes y)=y \otimes x$, for $x\in \mathbb{C}^{d_1}$, $y\in \mathbb{C}^{d_2}$. One obtains then the relations
\begin{equation}\label{eq:RelationPartialTracesFlip}
    \tr_1[C]=\tr_2[FCF^*] \quad \text{ and } \quad \tr_2[C]=\tr_1[FCF^*].\footnote{Here $FCF^*$ is understood on a swapped tensor product $\mathbb{C}^{d_2} \otimes \mathbb{C}^{d_1}$.}
\end{equation}

Given a self-adjoint matrix $C \in \L{d_1, d_2}$ with  spectrum $\lambda(C) \in \mathbb{R}^{d_1d_2}$, we will sometimes denote
\begin{equation*}
    \lambdaone(C):=\lambda(\tr_1 [C]) \in \mathbb{R}^{d_2} \, , \quad \lambdatwo(C):=\lambda(\tr_2 [C]) \in \mathbb{R}^{d_1}\, , \quad \lambdasum{C} \coloneqq \lambdaone(C) \oplus \lambdatwo(C)\, ,
\end{equation*}
 for the sake of brevity.

\subsection{Majorization theory}

For two vectors $x,y \in \mathbb{R}^d$,  there are different notions of majorization. If we denote by $x^{\downarrow},y^{\downarrow}$ the vectors containing the elements of $x,y$ rearranged in the decreasing order, one can define:
\begin{itemize}
    \item[1.] The elementwise majorization $x \leq y$, by the condition $x_i^{\downarrow}\leq y_i^{\downarrow}$, for every $1\leq i \leq d$.
    \item[2.] The weak majorization $x \preceq_{\omega} y$, by
    \begin{equation*}
        \sum_{i=1}^k x_i^{\downarrow} \leq \sum_{i=1}^k y_i^{\downarrow},
    \end{equation*}
    for every $1 \leq k \leq d$.
    \item[3.] Majorization $x \preceq y$, by $x \preceq_{\omega} y$ and also
    \begin{equation*}
        \sum_{i=1}^d x_i=\sum_{i=1}^d y_i.
    \end{equation*}
\end{itemize}
When we check whether  $x \preceq_{\omega} y$
 we  require that for every 
$k$, the sum of the 
$k$ largest components of $x$ is smaller or equal to  the sum of the 
$ k$ largest components of 
$y$. 
At first glance, this might suggest that it is necessary that $x, y$ are explicitly written in the decreasing order. However, checking majorization does not require knowing the order of entries in the vectors. Instead, one can check that for each $k\leq d$ any sum of $k$ entries of $x$ is upper bounded by some sum of $k$ entries of $y$.

\begin{definition}[\cite{Ando.1989}]
    A map $f:\text{dom}(f)\subseteq \mathbb{R}^n \to \mathbb{R}^m$ is said to be monotone increasing if it is order preserving w.r.t. $\leq$, i.e.
    \begin{equation*}
        f(x)\leq f(y)\quad \text{whenever} \quad x\leq y.
    \end{equation*}
    Similarly, $f$ is monotone decreasing if
        \begin{equation*}
        f(x)\leq f(y)\quad \text{whenever} \quad x\geq y.
    \end{equation*}
\end{definition}

\begin{definition}[\cite{Ando.1989}]
    A map $f:\text{dom}(f)\subseteq \mathbb{R}^n\to \mathbb{R}$ is called Schur-convex if it is order preserving w.r.t. $\preceq$, i.e.
\begin{equation*}
        f(x)\leq f(y)\quad \text{whenever} \quad x\preceq y.
    \end{equation*}
and  $f$ is Schur-concave if 
\begin{equation*}
        f(x)\leq f(y)\quad \text{whenever} \quad x\succeq y.
    \end{equation*}
\end{definition}
Given a  matrix $C\in \L{n}$ with eigenvalues $\{\lambda_i\}_{i=1}^n$ and singular values $\{\sigma_i\}_{i=1}^n$, we can consider Schur-convex or Schur-concave functions acting on the eigenvalues or singular  values of $C$, and this allows us to extend the definition to matrix spaces. We will call these extensions Schur-convex or Schur-concave functionals.

Examples of Schur-convex functionals are the Schatten $p$-norms, the Ky Fan $k$-norms or the operator norm that corresponds to the Ky Fan $1$-norm. On the contrary, the most well-known Schur-concave functionals are the $p$-quasinorms for $0<p<1$ or the minimum eigenvalue functional; for positive semidefinite matrices the determinant, and for a quantum state $\rho$ (a positive semidefinite matrix with trace one) the von Neumann entropy and the  Rényi $\alpha$-entropies for $\alpha \in (0,1)\cup (1,\infty)$. 
Notice also that $f$ is Schur-convex if and only if $-f$ is Schur-concave.  In \cref{Table:Schur} we provide the definitions for these quantities and the Schur-convexity properties that can be found  in \cite[Chapter 3]{Marshall.2011} and \cite[Chapter 13, F.3. (p. 562)]{Marshall.2011}.

\begin{table}
\centering
\scalebox{0.9}{%
\begin{tabular}{|c|c|c|c|}
\hline
Functional & Definition & Function & Convexity \\ \hline
Schatten $p$-norms & $\displaystyle \Vert C \Vert_p =\left(\sum_{i=1}^n\sigma_i^p \right)^{1/p}$ & $f(x)=\displaystyle\left(\sum_{i=1}^n x_i^p\right)^{1/p}$, $p \geq 1$ & Schur-convex  \\ \hline
Ky Fan norms & $\displaystyle \Vert C \Vert_{(k)}=\sum_{i=1}^k \sigma_i$ & $f_k(x)=\displaystyle\max_{i_1<\hdots<i_k}  x_{i_1} + \hdots +  x_{i_k}$ & Schur-convex  \\ \hline
Operator Norm & $\Vert C \Vert_{\infty}=\sigma_1$ & $f(x)=\max(x)$ & Schur-convex  \\ \hline
Minimum singular value &  $\sigma_n$ & $f(x)=\min(x)$ & Schur-concave \\ \hline
von Neumann entropy & $\displaystyle S(\rho)=-\sum_{i=1}^n \lambda_i \ln \lambda_i$ & $\displaystyle f(q)=-\sum_{i=1}^n q_i \ln q_i$ & Schur-concave  \\ \hline
Rényi $\alpha$ entropy &  $ \displaystyle  S_{\alpha}(\rho)=\frac{1}{1-\alpha}\ln\left(\sum_{i=1}^{n} \lambda_i^{\alpha} \right)$& $\displaystyle  f(q)=\frac{1}{1-\alpha}\ln\left(\sum_{i=1}^{n} q_i^{\alpha} \right)$ & Schur-concave \\ \hline
Determinant& $\displaystyle\det C=\prod_{i=1}^n \lambda_i$, $C >0$ & $f(x)=\displaystyle \prod_{i=1}^n x_i$, $x>0$ &  Schur-concave \\ \hline
\end{tabular}
}
\caption{Summary of main Schur-convex and Schur-concave functionals. The vectors the functions act on are $x,q \in \mathbb{R}^n$, where $x \geq 0$ and $q$ is a probability distribution.}
\label{Table:Schur}
\end{table}

\begin{rem}\label{rem:a<b=>ac<bc}
An observation which will play an important role in this work is that whenever $x \preceq y$, concatenating any vector $z$ to both $x$ and $y$, is an operation that strongly preserves majorization, i.e. 
$$x \preceq y \text{ if and only if} \begin{pmatrix}
        x \\ z
    \end{pmatrix} \preceq \begin{pmatrix}
        y \\ z
    \end{pmatrix}.$$
This  majorization result follows, for example, from a characterization of vector majorization via continuous convex functions, see \cite[Theorem I.4.B.1]{Marshall.2011}.
 \end{rem}

Finally, we will frequently make use of Schur's   majorization theorem, which states that the diagonal entries of any self-adjoint matrix $C \in L(\mathbb{C}^d)$ are majorized by the eigenvalues of $C$, see \cite[Theorem II.9.B.1.]{Marshall.2011}. That is, \begin{equation}\label{eq:Schur}
    \diag(C) \preceq \lambda(C)\end{equation}
The converse is also true: Horn's theorem \cite[Theorem II.9.B.2]{Marshall.2011} states that for any $c, \lambda \in \RRR^d$ with $c\preceq \lambda$ there exists a symmetric matrix $C$ with diagonal $c$ and spectrum $\lambda$.

\begin{rem}\label{remark:Decreasing order}
    Whenever we deal with diagonal matrices, we will denote by $\Lambda$  the diagonal matrix of eigenvalues of a matrix and by $\Sigma$ the diagonal matrix of singular values of a matrix. We will always assume (unless stated otherwise) that the diagonal elements are ordered in the decreasing order, that is $\Lambda=\Lambda^{\downarrow}$ and $\Sigma=\Sigma^{\downarrow}$. 
\end{rem}

\section{Local diagonalization of partial traces and majorization}\label{sec:DiagonalizationPartialTraces}

The  key idea underlying this work is  the observation that the eigenvalues of partial traces  arise from diagonal entries of a self-adjoint matrix when looking on the appropriate global basis.  This is the spirit of the  following result that will play a fundamental role throughout this work.

\begin{lemma}[Local Diagonalization Lemma]\label{lem:C_hat}
    Let $C\in\L{d_1,d_2}$  be a self-adjoint matrix. Then there exists a unitarily similar matrix $\hat{C}$  of $C$ such that:
    \begin{itemize}
        \item[(i)] $\lambda(C)=\lambda(\hat{C})$.
        \item[(ii)]$\lambda(\tr_i [C])=\lambda(\tr_i [\hat{C}])$, for $i=1,2$.
        \item [(iii)] The partial traces of $\hat{C}$ are diagonal.
        \item[(iv)] The eigenvalues of the partial traces of $C$ are given by \begin{equation}\label{eq:lambda:one,two}
\lambda^{(1)}_{i}(C)=\sum_{s=1}^{d_1}\hat c_{(s-1)d_2+i}, \quad i=1,\dots,d_2,
\qquad \lambda^{(2)}_{j}(C)=\sum_{s=1}^{d_2}\hat c_{(j-1)d_2+s}, \quad j=1,\dots,d_1\,
\end{equation}
where $\hat{c}_i$ denotes the $i$-th diagonal entry of $\hat{C}$.
    \end{itemize}
    
\end{lemma}
\begin{proof}
    If $C$ is a self-adjoint matrix, then both partial traces $\tr_1 [C]$ and $\tr_2 [C]$ are also self-adjoint. Consider  unitaries $U_1 \in \U{d_2}$ and $U_2 \in \U{d_1}$ that  diagonalize $\tr_1 [C]$ and $\tr_2 [C]$ respectively and define $\hat{C}=(U_2 \otimes U_1)C(U_2 \otimes U_1)^*$, so condition (i) is clear. Properties (ii) and (iii) follow from
    \begin{equation}
        \tr_1 [\hat{C}]=U_1 (\tr_1 [C] ) U_1^* \, , \qquad \tr_2 [\hat{C}]=U_2 (\tr_2 [C]) U_2^* \, .
    \end{equation}
    Therefore, the spectrum of the partial traces of $C$ can be read from the diagonal entries of the partial traces of $\hat{C}$ and thus, \eqref{eq:lambda:one,two} follows from \eqref{eq:diag:one,two}. 
\end{proof}

\begin{rem}
    Throughout the paper, we shall always choose  $\hat{C}$, i.e. unitary matrices $U_1$ and $U_2$, such that the vectors
$\lambda^{(1)}(\hat C)$ and $\lambda^{(2)}(\hat C)$ are listed in the decreasing order. 
\end{rem}

We now go one step further and look at the behavior under majorization of diagonals and eigenvalues of partial traces over unitary orbits.  Since partial traces preserve self-adjointness, we can apply Schur's theorem \eqref{eq:Schur} to obtain that $\diag(\tr_i [C]) \preceq \lambda(\tr_i [C])$, for $i=1,2$. On the other hand,  \cref{lem:C_hat} shows that for any matrix $C$ there exists a matrix $\hat{C}$ on its unitary orbit such that the eigenvalues of the partial traces of $C$ coincide with the  diagonal of the partial traces of $\hat{C}$. Therefore,  for each $i \in \{1,2\}$ the majorization structure of the sets\[\{\lambda(\tr_i [U C U^*]) \ : \ U \in \U{d_1, d_2}\} \quad \text{ and } \quad \{\diag(\tr_i [U C U^*]) \ : \ U \in \U{d_1, d_2}\}\] is the same, meaning that any element in one set is majorized by some element of the other set\footnote{This corresponds to the notion of majorization for matrix classes introduced in \cite{DahlGutermanS}.}. More specifically, we obtain the following result.

\begin{proposition}\label{prop:diagonalization}
    Let  $C \in \L{d_1,d_2}$ be a self-adjoint matrix. Then for $i = 1,2$,
    \begin{equation*}
        \diag(\tr_i [ C ]) \preceq \lambda(\tr_i [ C ])\, ,
    \end{equation*}
    and there exists $U \in  \U{d_1, d_2}$ such that
    \begin{equation*}
        \lambda(\tr_i [ C ])=\diag(\tr_i [U C U^*])\, . 
    \end{equation*}
\end{proposition}

\medskip

In particular, in  the proposition above, we conclude that for every Schur-convex function $f$, \begin{equation}\max\limits_{U \in \U{d_1,d_2}}f(\diag(\tr_i[UCU^*]))=\max\limits_{U \in \U{d_1,d_2}}f(\lambda(\tr_i[UCU^*]))\end{equation}
and
\begin{equation}\max\limits_{U \in \U{d_1,d_2}}f(\diag(\trsum{[UCU^*]}))=\max\limits_{U \in \U{d_1,d_2}}f(\lambda(\trsum{[UCU^*]})),\end{equation}
provided that $f$ is defined for all diagonals and spectra above.

\section{Optimal bounds for a single partial trace}\label{sec:SinglePartialTrace}

In this section, given a matrix $C\in\L{d_1,d_2}$, our first  goal is to analyze the maximum value over unitary orbits of functions depending on the partial traces of $C$, namely
\begin{equation}\label{eq:MajorizationProblemUnitaryOrbits}
    \max_{U \in \U{d_1,d_2}}f(\tr_i[UCU^*])\, , \quad i=1,2 \, .
\end{equation}
We explicitly compute this maximum in the case where $C$ is self-adjoint and 
$f$ is a Schur-convex functional, by employing majorization techniques applied to the eigenvalues of the corresponding partial traces. Subsequently, we extend our analysis to compute
\begin{equation}\label{eq:MajorizationProblemUnitaryOrbitsSVD}
    \max_{U,V \in \U{d_1,d_2}}f(\tr_i[UCV])\, , \quad i=1,2 \, .
\end{equation}
for any matrix $C \in \L{d_1,d_2}$ and  monotonically increasing Schur-convex functional $f$.

After the study of the maxima we will also study the minima of Schur-convex functionals over unitary orbits for the particular case $d_1=d_2$. We will see that in this scenario, the minimum for positive semidefinite matrices is achieved in the Bell basis. Recall that we always assume that $\Lambda=\Lambda^{\downarrow}$ and $\Sigma=\Sigma^{\downarrow}$ (see \cref{remark:Decreasing order}).

\subsection{Maxima}

In order to compute the maxima \eqref{eq:MajorizationProblemUnitaryOrbits} and \eqref{eq:MajorizationProblemUnitaryOrbitsSVD} we make use of the basis constructed in \cref{lem:C_hat} that preserves the spectra of  partial traces and also allows us to compute  eigenvalues and singular values in terms of the diagonal entries of the transformed matrix.

\begin{proposition}\label{MajorizationPartialTrace}
For any self-adjoint matrix $C \in \L{d_1, d_2}$,
    \begin{equation*}
      \lambda(\tr_2 [C] )\preceq\lambda(\tr_2 [\Lambda]), \quad \lambda(\tr_1 [C] )\preceq\lambda(\tr_1 [F^*\Lambda F])\, .
    \end{equation*}
     Consequently, for all Schur-convex functionals $f_i$, $i=1,2$,
    \begin{equation}
        \max_{U \in \U{d_1,d_2}}f_2(\tr_2 [UCU^*])= f_2(\tr_2 [\Lambda]) \, , \quad  \max_{U \in \U{d_1,d_2}}f_1(\tr_1 [UCU^*])= f_1(\tr_1 [F^*\Lambda F])\, ,
    \end{equation}
    provided that the eigenvalues of the partial traces lie within the domain of the functions.
\end{proposition}
\begin{proof}
    Let $C \in \L{d_1, d_2}$ be a self-adjoint matrix with spectrum $\lambda(C)$. Then by Lemma \ref{lem:C_hat} there exists a matrix $\hat{C}$ with $\lambda(\hat{C})=\lambda(C)$ such that for every $ i \in \{1,\hdots,d_1\}$,
    \begin{equation*}
        \lambda_i^{(2)}(C)=\sum_{s=1}^{d_2}\hat{c}_{(i-1)d_2+s}.
    \end{equation*}
    By Schur's theorem \eqref{eq:Schur}, $\diag(\hat{C})\preceq \lambda(C)$,  and we obtain that for every $1\leq k \leq d_1$
    \begin{equation*}
        \sum_{i=1}^{k} \lambda_i^{(2)}(C)=\sum_{i=1}^{k}\sum_{s=1}^{d_2}\hat{c}_{(i-1)d_2+s}\leq  \sum_{i=1}^{kd_2} \lambda_i =  \sum_{i=1}^{k} \lambda_i^{(2)}(\Lambda
        )\, .
    \end{equation*}
    The equality condition for $k=d_1$ follows from the trace preservation  of $\tr_2$. 
   
   Finally, the result for the first partial trace follows by applying  the already proved result to $FCF^* \in \L{d_2, d_1}$ and applying the flip relation \eqref{eq:RelationPartialTracesFlip}:
   \begin{equation}
       \lambda(\tr_1 [C] ) = \lambda(\tr_2[FCF^*]) \preceq \lambda(\tr_2 [\Lambda]) =\lambda(\tr_1 [F^*\Lambda F]),
   \end{equation}
   where the two middle partial traces are understood on the swapped tensor structure $\mathbb{C}^{d_2} \otimes \mathbb{C}^{d_1}$.
\end{proof}

\begin{corollary}\label{coro:PartialTraceAnyNorm}
    Let $C \in \L{d_1, d_2}$ be a self-adjoint matrix. Then, for any unitarily invariant norm $\unorm{\hspace{2pt} \cdot \hspace{2pt}}$,
    \begin{equation}
        \unorm{\tr_2 [C]}  \leq \unorm{\tr_2 [\Lambda]}, \quad \text{ and }\quad  \unorm{\tr_1 [C]}  \leq \unorm{ \tr_1 [F^*\Lambda F]}\, .
    \end{equation}
\end{corollary}
\begin{proof}
    Let $k \geq 1$. Since $\lambda(\tr_2 [C])\preceq \lambda(\tr_2 [\Lambda])$ by Proposition \ref{MajorizationPartialTrace}, we obtain  that $\Vert \tr_2[ C] \Vert_{(k)}\leq \Vert \tr_2 [\Lambda]\Vert_{(k)}$. The result follows then by Fan's dominance theorem \cite[Theorem IV 2.2]{Bhatia.1997}. Analogously, the same result holds for the first partial trace.
\end{proof}

\begin{theorem}\label{theo:WeakMajorizationSingularValues}
For every $C \in L(\mathbb{C}^{d_1} \otimes \mathbb{C}^{d_2})$,
\begin{equation}
    \sigma(\tr_2 [C])\preceq_{\omega} \sigma(\tr_2 [\Sigma]), \quad \sigma(\tr_1 [C])\preceq_{\omega} \sigma(\tr_1 [F^*\Sigma F]) \, ,
\end{equation}
where $\Sigma=\Sigma^{\downarrow}$
\end{theorem}
\begin{proof}
    We first note that for every $C$ there exists a self-adjoint $\hat{C}$ such that $\sigma(\tr_2 [C])=\sigma(\tr_2 [\hat{C}])$. We can see this easily by writing the singular value matrix $\Sigma_2$ of $\tr_2[C]$, $$ \Sigma_2 = U_2 \tr_2 [C] V_2 = \tr_2[(U_2 \otimes \1) \, C \, (V_2 \otimes \1)] = \tr_2[C_2]\, ,$$ where $C_2=(U_2 \otimes \1) \, C \, (V_2 \otimes \1)$. Note that now both sides are self-adjoint. This in consequence gives $\tr_2[C_2] = \tr_2[C_2^*]$ and therefore 
    \begin{equation*}
        \tr_2[C_2] = \tr_2\left[\frac{1}{2}(C_2 +  C_2^*) \right] + \tr_2\left[\frac{1}{2}( C_2 -  C_2^*)\right] = \tr_2\left[\frac{1}{2}( C_2 +  C_2^*)\right] = \tr_2 [\hat C] \, ,
    \end{equation*}
    where $\hat{C}=\frac{1}{2}( C_2 +  C_2^*)$ and it is  self-adjoint by construction. Now, we use that in Löwner order $\hat C \le |\hat C|$ and since the partial trace is a positive  map also $0\le \Sigma_2 \le \tr_2 [\hat C] \le \tr_2 [|\hat C|]$. In other words, $\vert \tr_2 \hat{C} \vert \leq \tr_2 \vert \hat{C}\vert $. This in particular gives us that $\sigma(\tr_2 [\hat C])\preceq_{\omega}\sigma(\tr_2 [\vert \hat C\vert] )$ by \cite[Proposition II.9.L.1]{Marshall.2011}.
    
    Finally, we have that  the singular values of $\hat{C}$ are weakly majorized by the ones of $C_2$, which follows from the explicit construction of $\hat{C}$  and Ky Fan singular value inequality (\cite[Theorem 5]{Fan.1951}).  Since $\sigma(C)=\sigma(C_2)$, we obtain that $\sigma(\hat{C})\preceq_{\omega} \sigma(C)$, which in turn  implies $\sigma(\tr_2[\hat{\Sigma}])\preceq_{\omega} \sigma(\tr_2 [\Sigma])$, where $\hat{\Sigma}$ is the singular value matrix of $\hat{C}$. The result follows then by the following chain  of inequalities
    \begin{equation*}
        \sigma(\tr_2 [C])=\sigma(\tr_2 [\hat{C}]) \preceq_{\omega} \sigma(\tr_2 [\vert \hat{C}\vert])\preceq \sigma(\tr_2 [\hat{\Sigma}]) \preceq_{\omega}\sigma(\tr_2 [\Sigma]).
    \end{equation*}
    where $\sigma(\tr_2 [\vert \hat{C}\vert])\preceq \sigma(\tr_2 [\hat{\Sigma}])$   follows from \cref{MajorizationPartialTrace}, since  $\hat{\Sigma}$ is the diagonal matrix of the positive semidefinite matrix $\vert \hat{C}\vert$. Finally, the result for $\tr_1$ follows from \eqref{eq:RelationPartialTracesFlip}.
\end{proof}

\begin{rem}[Monotonicity and weak majorization]
    In contrast to the case of self-adjoint matrices, for general matrices weak majorization alone does not allow the application of  Schur-convex functionals. In addition, the condition that the function is monotonically increasing must be imposed (see \cite[Theorem I.3.A.8]{Marshall.2011}). Functions like the Schatten $p$-norm ($p \geq 1$) are monotonically increasing but only on vectors in $\mathbb{R}^d_+$. Therefore, we can apply the previous result to the positive semidefinite matrix of the polar decomposition of $\tr_2 [C]$, which has the same singular values as $\tr_2 [C]$, and obtain
    \begin{equation}\label{ineq:PartialTracePNorm}
        \Vert \tr_2 [C] \Vert_p \leq \Vert \tr_2 [\Sigma] \Vert_p.
    \end{equation}

\end{rem}

\begin{rem}[Schatten Norms]
    We now turn our attention to  inequality \eqref{ineq:PartialTracePNorm} in more detail. The previous upper bounds for the $p$-norm involving partial traces of a general matrix were studied in \cite{Rastegin.2012,rico2025new,rico2025partial}, where it is shown that
    \begin{equation}\label{ineq:rastegin}
        \Vert \tr_2 [C] \Vert_p\leq d_2^{\frac{p-1}{p}}\Vert C \Vert_p, \quad  \Vert \tr_2 [C] \Vert_p\leq r^{\frac{p-1}{p}}\Vert C \Vert_p \, ,
    \end{equation}
   where $r$ is the rank. Notice that   the right-hand sides of the inequalities depend uniquely on the singular values of $C$. However, these bounds might not be sharp for every  vector of singular values $\sigma$. For example, consider the case where $d_1=d_2=2$, $p=2$ and $C$ a  matrix with non-zero singular values $\sigma_1>\hdots> \sigma_4>0$. Then
    \begin{equation*}
        \Vert \tr_2 [\Sigma] \Vert_2^2=(\sigma_1+\sigma_2)^2+(\sigma_3+\sigma_4)^2 < 2(\sigma_1^2+\sigma_2^2+\sigma_3^2+\sigma_4^2)=2 \Vert C \Vert_2^2\, .
    \end{equation*}
   This presents then an improvement on the previous inequalities, which is  sharp in terms of the singular values.
    
   For self-adjoint matrices,  the optimal bound in terms of the eigenvalues is given by
   \begin{equation}
       \Vert \tr_2 [C] \Vert_p\leq \Vert \tr_2 [\Lambda]\Vert_p,
   \end{equation}
   for $p\geq 1$.
\end{rem}

\begin{rem}[Ky Fan Norms]
    For Ky Fan norms, $\Vert \hspace{2pt}\cdot \hspace{2pt} \Vert_{(k)}$ it was recently shown in \cite{rico2025partial} that 
    \begin{equation}\label{eq:KyFanBoundPartialTrace}
        \Vert \tr_2 [C] \Vert_{(k)}\leq \max\left\{ 1, \frac{r}{k} \right\}\Vert C \Vert_{(k)}\, .
    \end{equation}
    In a similar way to the Schatten norms, Ky Fan norms are also monotonically increasing, which combined with Theorem \ref{theo:WeakMajorizationSingularValues} shows that 
    \begin{equation}
        \Vert \tr_2 [C] \Vert_{(k)}\leq \Vert \tr_2 [\Sigma] \Vert_{(k)},
    \end{equation}
    for every $1 \leq k\leq d_1$ and for any matrix $C$. This allows us to improve \eqref{eq:KyFanBoundPartialTrace} as the following example shows: Let $C$ be positive semidefinite and assume that $d_2k < r$, then
    \begin{equation*}
        \Vert \tr_2 [C] \Vert_{(k)}\leq \Vert \tr_2 [\Lambda] \Vert_{(k)}=\sum_{i=1}^{d_2 k}\lambda_i \leq d_2\sum_{i=1}^k \lambda_i <\frac{r}{k}\Vert C \Vert_{(k)}\, .
    \end{equation*}
\end{rem}

\subsection{Minima}\label{subsec:Minimums}

Once we have results on the maxima, the next step is to look at the minima and we will assume that $d_1=d_2=d$. This problem was studied for the mutual information in \cite{jevtic-2012}, where it was shown that the minimum is achieved in the basis of generalized Bell states (see \cite{Bennett.1993}) given by
\begin{equation*}
    \psi_{nm}=\frac{1}{\sqrt{d}}\sum_{j=0}^{d-1}e^{2\pi i j n/d}\vert j \rangle \otimes \vert (j+m) \text{ mod } d \rangle \, .
\end{equation*}
For a normal matrix $C$, let $U_0$ be  the unitary matrix that transforms the diagonal basis $\{v_i\}_{i}$  of $C$ into the Bell basis. If we relabel the eigenvalues and eigenvectors with the indices $0\leq n,m \leq d-1$, then we can write
\begin{equation*}
    C_{\text{Bell}}=U_0 CU_0^*=\sum_{n,m=0}^{d-1}\lambda_{nm}U_0\vert v_{nm} \rangle \langle v_{nm}\vert  U_0^*=\sum_{n,m=0}^{d-1}\lambda_{nm}\vert \psi_{nm}\rangle \langle \psi_{nm}\vert.
\end{equation*} 
In \cite{jevtic-2012} it was shown that the partial trace of $C_{\text{Bell}}$ is proportional to the identity, and by the trace preserving property of the partial trace it must hold that 
\begin{equation}
     \tr_2[C_{\text{Bell}}]=\frac{\tr[C]}{d}I\, .
\end{equation}
If $C$ is positive semidefinite, then \cite[Lemma 2.3]{Zhang.1998} shows that the eigenvalues of $\tr_2[C_{\text{Bell}}]$ are majorized by the eigenvalues of $\tr_2(C)$. If we apply now any Schur-convex functional $f$, this gives us 
\begin{equation*}
    f(\tr_2 [C_{\text{Bell}}])\leq f(\tr_2 [C])\, .
\end{equation*}
Notice that the same basis is also the minimizer for the first partial trace. For quantum states, we can  then obtain optimal spectral bounds for Schur-convex or Schur-concave functionals acting on reduced quantum states.

\begin{corollary}
    For any quantum state $\rho \in 
    \operatorname{L}(\mathbb{C}^d \otimes \mathbb{C}^d)$, with diagonal matrix $\Lambda=\Lambda^{\downarrow}$, the following bounds are sharp in terms of the spectrum:
    \begin{itemize}
    \item For $p\geq 1$,  \begin{equation*}
    d^{\frac{1-p}{p}} \leq  \Vert \tr_2 [\rho] \Vert_p \leq \Vert \tr_2 [\Lambda]\Vert_p\, .
\end{equation*}
\item For $0<p < 1$,
\begin{equation*}
    \Vert \tr_2 [\Lambda] \Vert_p \leq \Vert \tr_2 [\rho] \Vert_p \leq  d^{\frac{1-p}{p}}\, .
\end{equation*}
\item For the von Neumann entropy,
\begin{equation*}
    S(\tr_2 [\Lambda]) \leq S(\tr_2 [\rho]  )\leq \ln d \, .
\end{equation*}
\item For the Rényi entropies with $\alpha  \in (0,1)\cup (1,\infty)$, 
\begin{equation*}
   S_{\alpha}(\tr_2 [\Lambda]) \leq S_{\alpha}(\tr_2 [\rho])\leq \ln d\, .
\end{equation*}
\item For the determinant,
\begin{equation*}
     \det(\tr_2 [\Lambda]) \leq \det( \tr_2 [\rho] ) \leq \left(\frac{1}{d}\right)^d \, .
\end{equation*}
\end{itemize}
\end{corollary}

 For quantum states, the upper bound for the sum of determinants in partial traces improves the bound given in \cite{Lin.2016} when both Hilbert spaces are isomorphic, $\cH_1 \cong\cH_2$, since
\begin{equation*}
    \det(\tr_1 [\rho])^d+ \det(\tr_2 [\rho])^d\leq \frac{2}{d^{d^2}}<1.
\end{equation*}
for $d\geq 2$.

\section{Sufficient conditions for simultaneous majorization of partial traces}\label{sec:sufficient2pt}

In the previous section, we computed the maxima \eqref{eq:MajorizationProblemUnitaryOrbits} for Schur-convex functionals $f$, showing that the maximum within the corresponding unitary orbit is attained at a diagonal representative. In this section, we turn to a more complex maximization problem that simultaneously involves both partial traces, namely
\begin{equation}\label{eq:ProblemMaximzerBothPartialTraces}
     \max_{U \in \U{d_1,d_2}}f\left(\tr_1[UCU^*]\oplus \tr_2[UCU^*]\right)\, .
\end{equation}
One might expect that a diagonal matrix within the unitary orbit of 
$C$ again provides the maximizer. However, we demonstrate that this is not always the case: additional spectral constraints are required for optimality.

Some instances of \eqref{eq:ProblemMaximzerBothPartialTraces} have been solved or partially addressed in the literature. For instance, Bravyi proved in \cite{Bravyi.2003}  that for a positive semidefinite matrix $C\in \L{2,2}$, the maximum of  \eqref{eq:ProblemMaximzerBothPartialTraces} is attained at a diagonal matrix  when  $f$ is minus the sum of the two smallest eigenvalues. Moreover, for the negative von Neumann entropy $-S$, another Schur-convex functional, it was shown in \cite{jevtic-CMI} that the maximum is attained at a diagonal matrix.

\subsection{Sufficient conditions for two copies of the same system}

We begin by analyzing the case $d_1=d_2=d$, since the case $d_1\neq d_2$ turns out to be more complex.\footnote{For example, adding a scalar matrix $\gamma I$ to $C$ adds a scalar matrix $d\gamma I$ to $\trsum{[C]}$. This operation preserves majorization, and in particular, removes the distinction between self-adjoint and positive semidefinite matrices. This is not the case when $d_1 \neq d_2$.} Our next result establishes that whenever $d\geq 6$ and the spectrum of $C$ has at most seven degrees of freedom and one eigenvalue with at least multiplicity $d^2-6$, the solution to \eqref{eq:ProblemMaximzerBothPartialTraces}  is indeed given by the diagonal matrix whose eigenvalues are arranged in  the decreasing order. We obtain similar results for smaller dimensions. An important observation before the proof can save a lot of computations.
\begin{rem}[Necessary constraints for the existence of partial traces]\label{rem:MajorizationPartialTracesByDiagLambda}

 Given a self-adjoint $C \in \L{d_1, d_2}$, \cref{lem:C_hat} shows that there exists a matrix $\hat{C}$ such that its eigenvalues coincide with those of $C$ and the eigenvalues  for both partial traces coincide too. Furthermore,  the eigenvalues can be computed using the diagonal entries of $\hat{C}$.

Notice that the expressions for $\lambda^{(1)}_i(C)$ and $\lambda^{(2)}_j(C)$ of \eqref{eq:lambda:one,two} share the entry $\hat c_{(j-1)d_2 + i}$. Consider arbitrary $k_1 \leq d_2$ and $k_2 \leq d_1$ and  consider the sum of the largest $k_1$ elements of $\lambda^{(1)}(C)$ and the largest $k_2$ elements of $\lambda^{(2)}(C)$. In this sum there are $k_1d_1 + k_2d_2$ diagonal entries of $\hat C$, from which exactly $k_1 k_2$ appear twice; we denote the set of their indices by $I$. The other $k_1d_1 + k_2d_2 - 2k_1k_2$  appear once; we denote the set of their indices by $J$. Then 
\begin{equation}\label{eq:estimated_by_two_sums_of_eigenvalues}
    \begin{split}
        \sum\limits_{i=1}^{k_1} \lambda^{(1)}_i(C) + \sum\limits_{j=1}^{k_2} \lambda^{(2)}_j(C) & = 2\sum\limits_{i \in I} \hat{c}_i + \sum\limits_{j \in J} \hat{c}_j \\
        & \leq \sum\limits_{i=1}^{k_1d_1 + k_2d_2 - k_1k_2} \lambda_i + \sum\limits_{j=1}^{k_1k_2} \lambda_j\, .
    \end{split}
\end{equation}
where in the last step we used that $\diag(\hat{C})\preceq \lambda(\Lambda)$ by Schur's majorization theorem \eqref{eq:Schur}. Similarly, one can also obtain  the lower bound for the  sum of the smallest entries:
\begin{equation}\label{eq:estimated_by_two_sums_of_eigenvaluesLowerBound}
    \begin{split}
        \sum\limits_{i=1}^{l_1} \lambda^{(1)}_{d_2-i+1}(C) + \sum\limits_{j=1}^{l_2} \lambda^{(2)}_{d_1-j+1}(C) & = 2\sum\limits_{i \in I} \hat{c}_i + \sum\limits_{j \in J} \hat{c}_j \\
        & \geq \sum\limits_{i=1}^{l_1d_1 + l_2d_2 - l_1l_2} \lambda_{d_1d_2-i+1} + \sum\limits_{j=1}^{l_1l_2} \lambda_{d_1d_2-j+1}\, .
    \end{split}
\end{equation}
\end{rem}

It is easy to see, that every individual inequality \eqref{eq:estimated_by_two_sums_of_eigenvalues} is tight.

\begin{rem}\label{rem:attained_at_a_diagonal}
    Every inequality \eqref{eq:estimated_by_two_sums_of_eigenvalues} is attained at a diagonal matrix in the unitary orbit of $C$. Indeed, one needs to place the $k_1k_2$ largest eigenvalues of $C$ in the positions indexed by $I$, the $k_1d_1 + k_2d_2 - 2k_1k_2$ largest eigenvalues out of the remaining ones in the positions indexed by $J$.
\end{rem}

For a permutation $\pi$ we will denote by $\Lambda_\pi$ the diagonal matrix with $(\Lambda_{\pi})_{ii}=\lambda_{\pi(i)}$.
One might be tempted to think that the remark above implies that for any $U \in \U{d_1, d_2}$ there exists a diagonal matrix $\Lambda_\pi$ in the unitary orbit of $C$ such that $\trsum{[UCU^*]}\preceq\trsum{\Lambda_\pi}$. While it is true that for every $k$ the sum of $k$ largest eigenvalues of $\trsum{[UCU^*]}$ is bounded by that of some $\trsum{[\Lambda_{\pi}]}$, different $k$ may require different $\Lambda_\pi$ and thus majorization can fail. However, majorization of the entire unitary orbit with one matrix can now be easily characterized.
\begin{corollary}
    Let $D\in \L{d_1, d_2}$ be a matrix in the unitary orbit of a self-adjoint matrix $C \in \L{d_1, d_2}$. Then the following are equivalent.
    \begin{itemize}
        \item[(i)] For any $U \in \U{d_1, d_2}$ \begin{equation}
        \lambdatrsum{UCU^*} \preceq \lambdatrsum{D}.
    \end{equation}
    \item[(ii)] For any diagonal matrix $\Lambda_\pi$ in the unitary orbit of $C$ \begin{equation}
        \lambdatrsum{\Lambda_\pi} \preceq \lambdatrsum{D}.
    \end{equation}
    \item[(iii)] for any $k \leq d_1 + d_2 - 1$ \begin{equation}\label{eq:max_k1_k2}
        \sum_{i=1}^{k}\lambda^\downarrow_i(\trsum{D}) = \max_{\substack{k_1\leq d_2,\\ k_2\leq d_1:\\ k_1 + k_2 = k}}\left[\sum\limits_{i=1}^{k_1d_1 + k_2d_2 - k_1k_2} \lambda_i + \sum\limits_{j=1}^{k_1k_2} \lambda_j\, \right].
    \end{equation}
    \end{itemize}
\end{corollary}
\begin{proof}
    Clearly (i) implies (ii). Note that by \cref{rem:attained_at_a_diagonal} \[\max_{\pi} \left[\sum_{i=1}^{k} \lambda^\downarrow_i(\trsum{\Lambda_\pi})\right] = \max_{\substack{k_1\leq d_2,\\ k_2\leq d_1:\\ k_1 + k_2 = k}}\left[\sum\limits_{i=1}^{k_1d_1 + k_2d_2 - k_1k_2} \lambda_i + \sum\limits_{j=1}^{k_1k_2} \lambda_j\, \right].\]
    
    If (ii) holds, then for any $k \leq d_1 + d_2 - 1$ \[\sum_{i=1}^{k} \lambda^\downarrow_i(\trsum{\Lambda_\pi}) \leq \sum_{i=1}^{k} \lambda^\downarrow_i(\trsum{D}).\] This shows that the right hand side of \eqref{eq:max_k1_k2} doesn't exceed the left hand side. The converse inequality follows from \eqref{eq:estimated_by_two_sums_of_eigenvalues}.

    Analogously, (iii) together with \eqref{eq:estimated_by_two_sums_of_eigenvalues} imply that for any $k \leq d_1 + d_2 - 1$ \[\sum_{i=1}^{k} \lambda^\downarrow_i(\trsum{UCU^*}) \leq \sum_{i=1}^{k} \lambda^\downarrow_i(\trsum{D}).\]
    Together with the trace preservation property this implies (i).
\end{proof}

\begin{rem}
    If $d_1 = d_2 \eqqcolon d$, then \begin{equation}\label{eq:k1k2ifd1=d2} \max_{\substack{k_1\leq d_2,\\k_2\leq d_1:\\ k_1 + k_2 = k}}\left[\sum\limits_{i=1}^{k_1d_1 + k_2d_2 - k_1k_2} \lambda_i + \sum\limits_{j=1}^{k_1k_2} \lambda_j\, \right] = \sum\limits_{i=1}^{dk - \lfloor\frac{k^2}{4}\rfloor} \lambda_i + \sum\limits_{j=1}^{\lfloor\frac{k^2}{4}\rfloor} \lambda_j\end{equation}
    because for a fixed $k=k_1+k_2$ the sum $\sum\limits_{i=1}^{kd - k_1k_2} \lambda_i + \sum\limits_{j=1}^{k_1k_2} \lambda_j$ is increasing in $k_1k_2$ and thus the maximum is attained at $k_1 = \lfloor\frac{k}{2}\rfloor$, $k_2 = \lceil\frac{k}{2}\rceil$.
\end{rem}

\begin{proposition}[Sufficient conditions]\label{prop:NewSufficientConditionsSquareCase}
    Let $d \geq 6$. Then, for any  self-adjoint $C \in  \L{d, d}$ with spectrum satisfying $\lambda_4=\hdots=\lambda_{d^2-3}$,
\begin{equation}\label{eq:NewjointMajorizationPartialTraces}
        \lambda(\tr_1 [C] \oplus \tr_2 [C])\preceq  \lambda(\tr_1 [\Lambda] \oplus \tr_2[\Lambda]) \, .
    \end{equation}
    Consequently, for every Schur-convex functional $f$,
    \begin{equation}
       \max_{U \in \U{d,d}} f(\tr_1 [UCU^*] \oplus \tr_2 [UCU^*])= f(\tr_1 [\Lambda] \oplus \tr_2[\Lambda])\, .
    \end{equation}
      provided that the eigenvalues of the partial traces lie within the domain of the functionals.
\end{proposition}
\begin{proof}
    Because $d_1 = d_2 = d$, the scalar shift does not affect \eqref{eq:NewjointMajorizationPartialTraces}. Therefore for our convenience we can assume that $\lambda_4=\hdots=\lambda_{d^2-3} = 0$ and \[\lambda(C)^T = (\lambda_1, \lambda_2, \lambda_3, 0, \ldots, 0, \nu_1, \nu_2, \nu_3),\]
    where $\lambda_1 \geq \lambda_2 \geq \lambda_3 \geq 0 \geq \nu_1 \geq \nu_2 \geq \nu_3$. Then
    \begin{equation*}
    \lambdasum{\Lambda}^T=\Bigg(\lambda_1 + \lambda_2 + \lambda_3, \lambda_1, \lambda_2, \lambda_3, \underbrace{0}_{2d-8},\nu_1, \nu_2, \nu_3, \nu_1 + \nu_2 + \nu_3\Bigg), \,
\end{equation*} which is rearranged in the decreasing order.

It remains to verify that for any $k \leq 2d - 1$ condition \eqref{eq:max_k1_k2} holds for $D = \Lambda$. One can verify this explicitly with Table  \ref{table:newSufficientConditions}, where both sides of the equation are provided.
\end{proof}
  \begin{table}
  \centering
$\begin{array}{| c | c | c |} 
 \hline
 k & \sum\limits_{i=1}^{k} \lambda^\downarrow_i(\trsum{\Lambda}) & \sum\limits_{i=1}^{dk - \lfloor\frac{k^2}{4}\rfloor} \lambda_i + \sum\limits_{j=1}^{\lfloor\frac{k^2}{4}\rfloor} \lambda_j\\
 \hline\hline
 1 & \lambda_1 + \lambda_2 + \lambda_3 & \sum\limits_{i=1}^{d} \lambda_i\\
 \hline
 2 & 2\lambda_1 + \lambda_2 + \lambda_3 & \sum\limits_{i=1}^{2d - 1} \lambda_i + \lambda_1\\
 \hline
 3 & 2\lambda_1 + 2\lambda_2 + \lambda_3 & \sum\limits_{i=1}^{3d - 2} \lambda_i + \sum\limits_{i=1}^{2}\lambda_j\\
 \hline
 4\leq k \leq 2d-4 & 2(\lambda_1 + \lambda_2 + \lambda_3) & \sum\limits_{i=1}^{4d - 4} \lambda_i + \sum\limits_{i=1}^{4}\lambda_j = \sum\limits_{i=1}^{d^2 - 4} \lambda_i + \sum\limits_{i=1}^{(d- 2)^2}\lambda_j\\
 \hline
 2d-3& 2(\lambda_1 + \lambda_2 + \lambda_3) + \nu_1 & \sum\limits_{i=1}^{d^2 - 2} \lambda_i + \sum\limits_{i=1}^{d^2 - 3d + 2}\lambda_j\\
 \hline
 2d-2& 2(\lambda_1 + \lambda_2 + \lambda_3) + \nu_1 + \nu_2 & \sum\limits_{i=1}^{d^2 - 1} \lambda_i + \sum\limits_{i=1}^{(d - 1)^2}\lambda_j\\
 \hline
 2d-1& 2(\lambda_1 + \lambda_2 + \lambda_3) + \nu_1 + \nu_2  + \nu_3 & \sum\limits_{i=1}^{d^2} \lambda_i + \sum\limits_{i=1}^{d^2 - d}\lambda_j\\
 \hline
\end{array}$
\caption{Equations for \cref{prop:NewSufficientConditionsSquareCase}.}
\label{table:newSufficientConditions}
\end{table}

 We can obtain a similar result for smaller dimensions. Notably, in case $d=2$ the majorization condition is always satisfied.
\begin{proposition}\label{prop:sufficient_conditions_small_dimensions}
    Let $3 \leq d \leq 5$ and $0 \leq p, q \leq 3$ with $p+q=d$. Then, for any  self-adjoint $C \in  \L{d, d}$ with spectrum satisfying $\lambda_{p+1}=\hdots=\lambda_{d^2-q}$,
\begin{equation}\label{eq:d<=5:NewjointMajorizationPartialTraces}
        \lambda(\tr_1 [C] \oplus \tr_2 [C])\preceq  \lambda(\tr_1 [\Lambda] \oplus \tr_2[\Lambda]) \, .
    \end{equation}
Additionally, Condition \eqref{eq:d<=5:NewjointMajorizationPartialTraces} always holds for $d=2$. In other words, the following spectral conditions are sufficient for \eqref{eq:d<=5:NewjointMajorizationPartialTraces}:
\begin{enumerate}
    \item $d = 2$;
    \item $d = 3$ and $\lambda_n=\hdots=\lambda_{n+5}$ for some $n \in \{1,2,3,4\}$,
    \item $d = 4$ and $\lambda_{n}=\hdots=\lambda_{n+11}$ for some $n \in \{2,3,4\}$;
    \item $d = 5$ and $\lambda_{n}=\hdots=\lambda_{n+19}$ for some $n \in \{3,4\}$;
\end{enumerate}
\end{proposition}
\begin{proof}
See \cref{propAPP:sufficient_conditions_small_dimensions} in the appendix.
\end{proof}

A family of matrices that always satisfy the assumption of Propositions \ref{prop:NewSufficientConditionsSquareCase} and \ref{prop:sufficient_conditions_small_dimensions} are the matrices of rank at most $3$. Consequently we obtain the following result:

\begin{corollary}\label{coro:SufficientConditionsd<4SquareCase}
    If $C \in \L{d,d}$ is self-adjoint and has rank $r\leq 3$, then
    \begin{equation*}
        \lambda(\tr_1 [C] \oplus \tr_2 [C])\preceq \lambda(\tr_1 [\Lambda] \oplus \tr_2 [\Lambda])\, .
    \end{equation*}
\end{corollary}
We will see in \cref{prop:sufficientRank3} that this result also holds for self-adjoint matrices $C \in \L{d_1,d_2}$ with rank $r\leq  3$, for every dimension $d_1,d_2 \geq 2$.

\begin{rem}[Applications to noisy quantum states]
For quantum states, this type of spectrum naturally arises in noisy families of low-rank states. For example, consider convex combinations of a low-rank state  $\sigma \in \L{d,d}$ with $\rk(\sigma)\leq 3$ and the maximally mixed state:
\begin{equation}
    \rho=p \sigma+\frac{1-p}{d^2}\1_{d^2}
\end{equation}
for $0\leq p \leq 1$. These states satisfy the conditions of Propositions \ref{prop:NewSufficientConditionsSquareCase}, \ref{prop:sufficient_conditions_small_dimensions}.  Such examples include depolarizing noise in a bipartite system of qubits, the  case where $\sigma$ is a general pure state, leading to noisy pure-state families such as isotropic states, as well as the case where $\sigma$ is a normalized projection onto a $2$- or $3$-dimensional space. For these states, we obtain inequalities such as
\begin{equation}
\Vert \tr_1[\rho] \Vert_p^p+\Vert \tr_2[\rho] \Vert_p^p\leq  \Vert \tr_1[\Lambda] \Vert_p^p+\Vert \tr_2[\Lambda] \Vert_p^p \, , \quad p \geq 1
\end{equation}
\begin{equation}
\det(\tr_1 [\rho])\det(\tr_2 [\rho])\geq  \det(\tr_1 [\Lambda])\det(\tr_2 [\Lambda])
\end{equation}
\begin{equation}
S(\tr_1[ \rho])+S(\tr_2[\rho])\geq S(\tr_1[ \Lambda])+S(\tr_2[\Lambda])\,,
\end{equation}
where  we  always assume $\Lambda=\Lambda^{\downarrow}$.
\end{rem}

\begin{rem}[Lower bound for the mutual information]\label{rem:LowerBoundMI}
    The mutual information is a measure of distinguishability between a bipartite quantum state $\rho_{AB}$ and the product of its marginals $\rho_A \otimes \rho_B$. It is defined as
    \begin{equation}
        I_{\rho}(A:B)=S(\rho_A)+S(\rho_B)-S(\rho_{AB}) \, ,
    \end{equation}
    and its value depends on the basis of the global system. A fundamental lower bound for this quantity is provided by Pinsker's inequality $I_{\rho}(A:B)\geq \frac{1}{2}\Vert \rho_{AB}-\rho_A \otimes \rho_B \Vert_1^2$. Moreover, it was observed in \cite[Sec. VI]{jevtic-CMI} that the minimum of $I_{\rho}(A:B)$ along unitary orbits is attained at  one of the diagonal matrices $\Lambda_{\pi}$ of $\rho_{AB}$, for some permutation of the eigenvalues $\pi$. However, there is no universal permutation that serves as the optimizer for arbitrary dimensions. Proposition \ref{prop:NewSufficientConditionsSquareCase} establishes that whenever the spectrum possesses at most seven degrees of freedom and one eigenvalue with multiplicity $d^2-6$, the permutation that minimizes the mutual information  is  the one that arranges the eigenvalues in  the decreasing  order.
\end{rem}

Having examined the spectral conditions sufficient to ensure that the maximum of \eqref{eq:ProblemMaximzerBothPartialTraces} is attained in a diagonal matrix, it is worth asking about the necessary conditions. In Section \ref{sec:BothPartialTracesSameDimension}, and particularly in Corollary \ref{coro:d>24} we will see that there is not much more margin to improve, given the rank of the matrix.

\subsection{Sufficient conditions for general dimensions}

In the case where the dimensions are equal, we saw in Proposition \ref{prop:NewSufficientConditionsSquareCase} that seven degrees of freedom in the spectrum and an eigenvalue with multiplicity $d^2-6$ is a sufficient condition for majorization \eqref{eq:NewjointMajorizationPartialTraces}. However,  we have to add more constraints on the eigenvalues. For example, if we demand that $\Lambda=F^*\Lambda F$, then the majorization follows immediately by \cref{MajorizationPartialTrace}. This is the case of the next result.

\begin{proposition}\label{prop:SufficientCaseDifferentDimensions}
    Let $2\leq d_1<d_2$ and let $C \in \L{d_1,d_2}$ be self-adjoint with spectrum of the form $\lambda(C)=(\lambda_1, \lambda, \hdots, \lambda, \lambda_{d_1d_2})$ sorted in the decreasing order. Then,
    \begin{equation*}
        \lambda(\tr_1 [C] \oplus \tr_2 [C])\preceq \lambda(\tr_1 [\Lambda] \oplus \tr_2 [\Lambda])\, .
    \end{equation*}
\end{proposition}

In an analogous way to \cref{coro:SufficientConditionsd<4SquareCase}, the statement  still remains true for matrices with rank at most three.

\begin{proposition}\label{prop:sufficientRank3}
    Let $2\leq d_1<d_2$ and $C \in \L{d_1,d_2}$ self-adjoint with $\rk{C}\leq 3$. Then,
    \begin{equation*}
        \lambda(\tr_1 [C] \oplus \tr_2 [C])\preceq \lambda(\tr_1 [\Lambda] \oplus \tr_2 [\Lambda])\, .
    \end{equation*}
\end{proposition}
\begin{proof}
      Assume first that  $C \geq 0$. In this case,
    \begin{equation*}
        \lambdasum{\Lambda}^T=(\lambda_1+\lambda_2+\lambda_3,\lambda_1,\lambda_2, \lambda_3,0,\hdots,0)
    \end{equation*}
     Consider arbitrary $1 \leq k < d_1+d_2$ and decompose it as $k=k_1+k_2$, with $k_1\leq d_2$, $k_2 \leq d_1$.  Following Remark \ref{rem:MajorizationPartialTracesByDiagLambda}, we need to check that 
\begin{equation*}\sum\limits_{i=1}^{k_1}\lambda^{(1)}_i + \sum\limits_{j=1}^{k_2}\lambda^{(2)}_j \leq \sum_{i=1}^k \lambda_i^{(1,2)}(\Lambda)\end{equation*} holds for every $k$. The cases $k=1$ and $k\geq 4$ are straightforward, and  for $2 \leq k \leq 3$ notice that we always have $k_1 k_2 \leq k - 1$.
Therefore, inequality \eqref{eq:estimated_by_two_sums_of_eigenvalues} implies 
     \begin{equation*}
\sum_{i=1}^{k_2}\lambda^{(1)}_i+\sum_{j=1}^{k_1}\lambda^{(2)}_j\leq \tr[C]+\sum_{i=1}^{k_1k_2}\lambda_i\leq (\lambda_1 + \lambda_2 + \lambda_3)+\sum_{i=1}^{k-1}\lambda_i \, ,
     \end{equation*}
     and the result holds. Notice that the result also holds when the three eigenvalues are negative by looking at $-C$. To conclude, it is enough to check the case where $C$ has eigenvalues $\lambda_1,\lambda_2>0$ and $\lambda_3<0$, since it also shows the case with one positive and two negative eigenvalues by looking at $-C$. However, for $\lambda_1,\lambda_2>0$ and $\lambda_3<0$, it can be easily checked that majorization holds by considering the matrix with eigenvalues in the decreasing order and 
    \begin{equation*}
        \lambda^{(1,2)}(\Lambda)^T=(\lambda_1+\lambda_2, \lambda_1,\lambda_2,0, \hdots,0, \lambda_3,\lambda_3)\, .
    \end{equation*}
\end{proof}

\section{Quadratic Programs for obtaining general bounds}\label{sec:QP}

In this work we devoted a great effort to finding sufficient conditions under which we could majorize the eigenvalues of partial traces with the respective eigenvalues of partial traces of a diagonal matrix. However, such a phenomenon occurs only in very particular cases as we saw in Propositions \ref{prop:NewSufficientConditionsSquareCase}, \ref{prop:sufficient_conditions_small_dimensions}, \ref{prop:SufficientCaseDifferentDimensions} or \ref{prop:sufficientRank3}. Therefore, in this section we develop tools to obtain upper bounds for general matrices for the problem \eqref{eq:ProblemMaximzerBothPartialTraces}, even though they might not be optimal anymore. 

To this end, let us consider first the case $d_1=d_2=d\geq 6$. Given a general self-adjoint matrix $C \in \L{d,d}$ we seek a vector $\mu$ that satisfies $\lambda(C)\preceq \mu$  and the sufficient condition of Proposition \ref{prop:NewSufficientConditionsSquareCase}:
% This sufficient conditions lead to the following four possible forms for $\mu$:
% \begin{equation}\label{eq:definitionmu1}
% \mu_I^T=(\mu_1, \mu_2, \mu_3, \mu_4, \hdots, \mu_4)\, , \quad    \mu_{II}^T=(\mu_1, \mu_2, \mu_3,\hdots, \mu_3,\mu_4)\, .
% \end{equation}
% \begin{equation}\label{eq:definitionmu2}
% \mu_{III}^T=(\mu_1, \mu_2,\hdots, \mu_2,\mu_3,\mu_4)\, , \quad \mu_{IV}^T=(\mu_1,\hdots , \mu_1\mu_2,\mu_3,\mu_4)\, .
% \end{equation}
% for $\mu_1\geq \mu_2 \geq \mu_3 \geq \mu_4$.
\begin{equation}\label{eq:definitionmu}
\mu^T=(\mu_1, \mu_2, \mu_3, \underbrace{\mu_4, \hdots, \mu_4}_{d^2-6}, \mu_{5}, \mu_{6}, \mu_{7})\ .
\end{equation}
Let $M=\diag(\mu^T)$. If we could bound 
\begin{equation}
      \lambdatrsum{[C]}\preceq \lambdatrsum{[M]}, 
\end{equation}
then we could apply Proposition \ref{prop:NewSufficientConditionsSquareCase} to the right-hand side and obtain bounds for the eigenvalues of partial traces of $C$. The next general result shows that this is indeed possible.

\begin{proposition}\label{prop:MajorizationM}
Let $C \in \L{d_1,d_2}$ be a self-adjoint matrix with spectrum $\lambda(C)$ and let $\mu  \in \mathbb{R}^{d_1d_2}$ satisfy $\lambda(C) \preceq \mu$.
Let $M= \diag{(\mu)}$ and let $\mathcal{M}$ be a subset of the unitary orbit of $M$.

If for any $U \in \U{d_1, d_2}$ there exists $D_U \in \mathcal{M}$ such that
\begin{equation}\label{eq:M<D}
    \lambdatrsum{[UMU^*]}\preceq \lambdatrsum{[D_U]}. 
\end{equation}

Then there exists $D \in \mathcal{M}$ such that
\begin{equation}\label{eq:C<D}
    \lambdatrsum{[C]}\preceq \lambdatrsum{[D]}. 
\end{equation}
\end{proposition}
\begin{proof}
   By \cref{lem:C_hat}, there exists a matrix $\hat{C}$ such that  $\lambda(\tr_1 [C] \oplus \tr_2 [C])= \diag(\tr_1 [\hat{C}]\oplus \tr_2 [\hat{C}])$. In addition, since  $\diag(\hat{C})\preceq \lambda(\hat{C})\preceq \mu$,  by Horn's theorem there exists $U \in \U{d_1, d_2}$ such that $\diag(UMU^*) = \diag(\hat{C})$. It follows that \[\diag(\trsum{[\hat{C}]}) = \diag(\trsum{[UMU^*]}) \preceq \lambda(\trsum{[UMU^*]})\, . \]
   Finally, \eqref{eq:C<D} follows from \eqref{eq:M<D}.
\end{proof}

In our case, for $\mu$ as in \eqref{eq:definitionmu} and $\mathcal{M}=\{\diag(\mu)\}$ the conditions of the previous Proposition are fulfilled. Moreover, the feasible set is described by \eqref{eq:definitionmu} and $\lambda \preceq \mu$, which is a collection of linear inequalities. The question now reduces to finding the best vectors $\mu$ that provide the best bounds. In our case, we will aim for $\mu$ to be the flattest admissible vector, since many Schur-convex functions minimize over these vectors as we already saw in \cref{subsec:Minimums}.

We will choose to minimize $\Vert \mu \Vert_2^2$ (since minimizing the $2$-norm square penalizes large components quadratically) under the additional constraint  $\lambda(C) \preceq \mu$. This minimization problem can be expressed in terms of a quadratic program (QP) with seven variables: $\mu_1,\mu_2, \ldots \mu_7 \in \mathbb{R}$:
 \begin{equation}\label{eq:SDPSameDimensions}
\begin{aligned}
& \underset{\mu\in \mathbb{R}^7}{\text{minimize}}
& & \mu_1^2+\mu_2^2+\mu_3^2+(d^2-6)\mu_4^2 + \mu_5^2 + \mu_6^2 + \mu_7^2 \\
& \text{subject to}
& &\mu_1\geq \mu_2\geq \mu_3\geq \mu_4 \geq \mu_5 \geq \mu_6 \geq \mu_7\\
& 
& &\lambda(C)\preceq \mu.\\
\end{aligned}
 \end{equation}

Analogously, one can slightly modify the QP \eqref{eq:SDPSameDimensions} to obtain QPs for the vectors $\mu$, corresponding to sufficient conditions in Proposition \ref{prop:sufficient_conditions_small_dimensions}. For example, when $d=3$, we obtain four possible forms for $\mu$:
\begin{equation}\label{eq:definitionmu1}
\mu_I^T=(\mu_1,\hdots , \mu_1,\mu_2,\mu_3,\mu_4)\, , \quad    \mu_{II}^T=(\mu_1, \mu_2,\hdots, \mu_2,\mu_3,\mu_4)\, .
\end{equation}
\begin{equation}\label{eq:definitionmu2}
\mu_{III}^T=(\mu_1, \mu_2, \mu_3,\hdots, \mu_3,\mu_4)\, , \quad \mu_{IV}^T=(\mu_1, \mu_2, \mu_3, \mu_4, \hdots, \mu_4)\, .
\end{equation}
for $\mu_1\geq \mu_2 \geq \mu_3 \geq \mu_4$.

A useful feature of this approach is that the optimization used to select $\mu$ need not be tailored to the Schur-convex functional of interest. Once an admissible $\mu$ is found, Proposition \ref{prop:MajorizationM} yields a majorization bound and therefore a bound for every Schur-convex functional simultaneously. Thus, the objective function serves only to select a favorable admissible majorant. We may therefore use the computationally convenient objective function $\Vert \mu \Vert_2^2$, obtaining a quadratic program, and then apply its solution even to Schur-convex functionals whose direct optimization is substantially more difficult.

\begin{example}\label{example:numerics}
    We now present an example where  our solutions of these QPs can improve the already existing bounds. Consider a density matrix $\rho \in L(\mathbb{C}^3 \otimes \mathbb{C}^3)$ with spectrum $$\lambda(\rho)^T=\frac{1}{45}(15,10,5,4,3,3,2,2,1)\, .$$ For the case of $p$-norms there are two possibilities in general, which are the bound obtained by Rastegin in \cite{Rastegin.2012}, which corresponds to the dimensional bound in \eqref{ineq:rastegin},  and  the bound obtained by Audenaert in \cite{Audenaert.2007}, which was later generalized to general matrices in \cite{rico2025partial}. Figure \ref{fig:pNorms} shows that $\mu_{II}$ \eqref{eq:definitionmu1} and $\mu_{III}, \mu_{IV}$ \eqref{eq:definitionmu2} improve the previous well-known bounds for this example. In particular, $\mu_{III}$ turns out to be the best bound among the others, and it also shows that $\Vert \tr_1 [\rho] \Vert_p^p+\Vert \tr_2 [\rho] \Vert_p^p \to 0$, when $p \to \infty$.
    %\begin{figure}
     %   \centering
      %  \includegraphics[width=0.7\textwidth]{plotnorms.png}
      %  \caption{Upper bounds for sum of $p$-norms of partial traces}
       % \label{fig:pNorms}
    %\end{figure}
    \begin{figure}[H]
    \centering
    \begin{minipage}{0.496\textwidth}
        \centering
        \includegraphics[width=\linewidth]{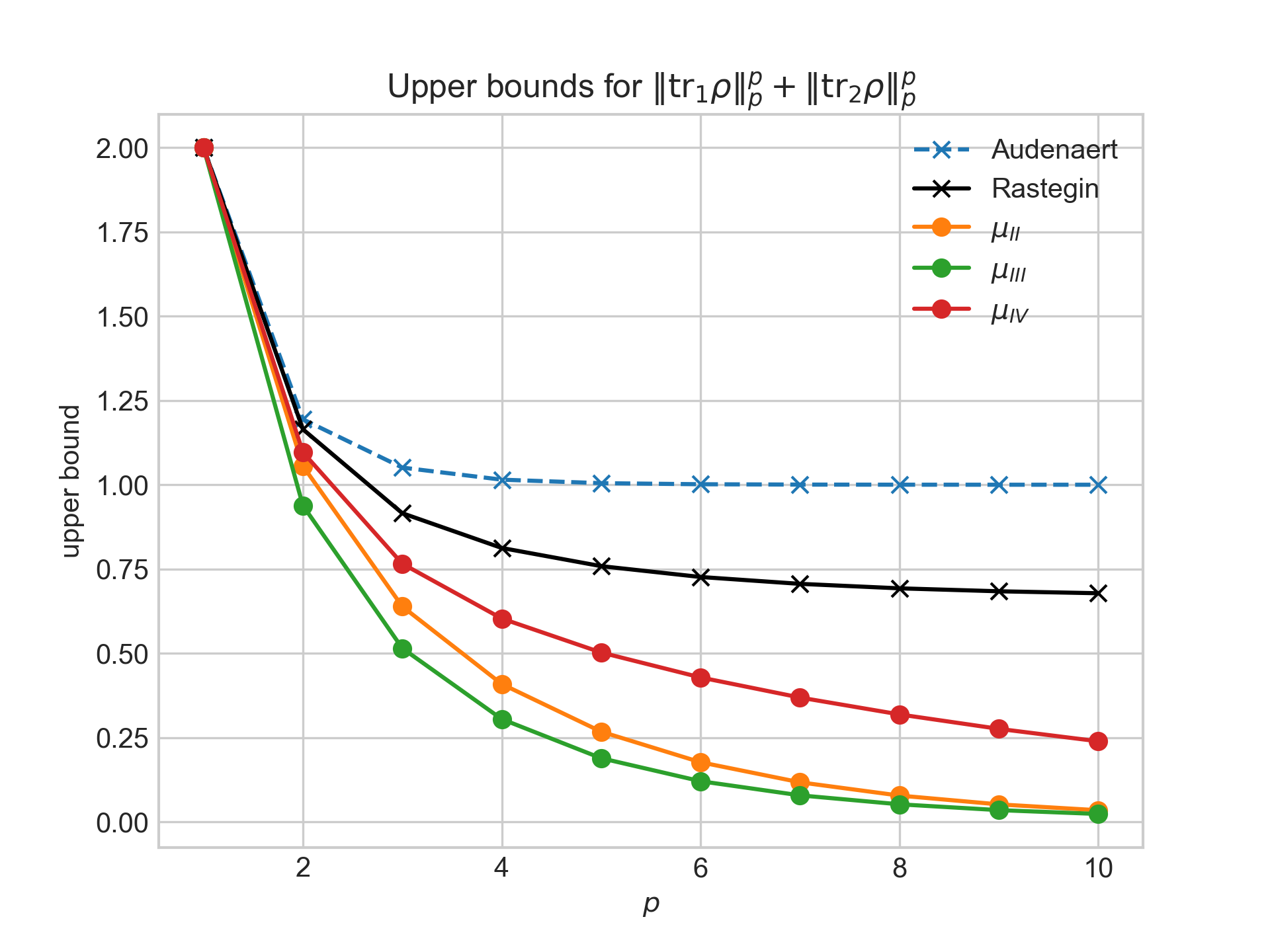}
        \caption{Upper bounds for sum of $p$-norms of partial traces}
         \label{fig:pNorms}
    \end{minipage}
    \hfill
    \begin{minipage}{0.496\textwidth}
        \centering
        \includegraphics[width=\linewidth]{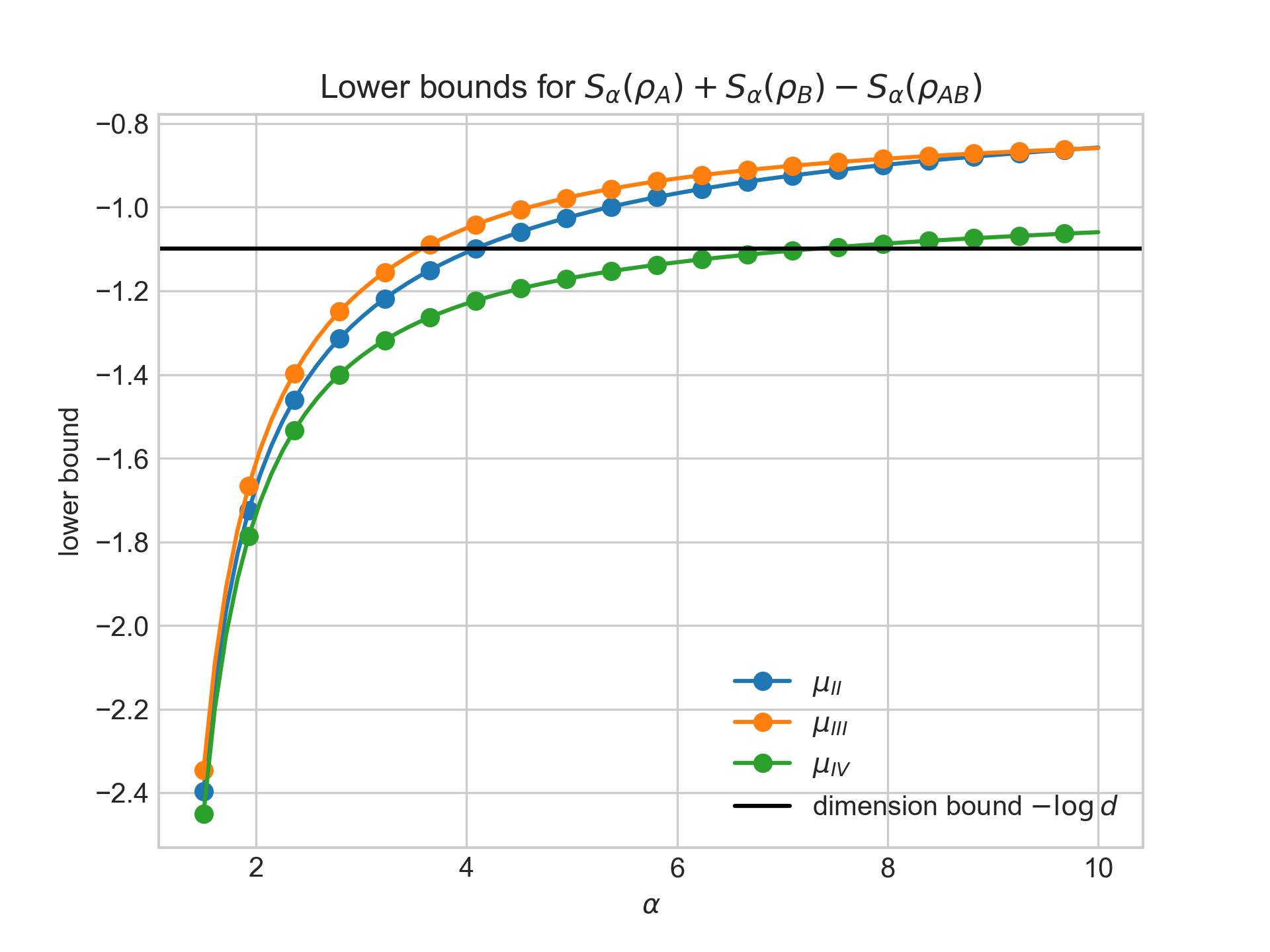}
        \caption{Lower bounds for subadditivity of Rényi entropies}
         \label{fig:AlphaRenyi}
    \end{minipage}
\end{figure}

    The reason $\mu_{I}$ is not included in Figure \ref{fig:pNorms} is that the bound is too weak due to negative eigenvalues on the partial traces, which are given as solutions of the QP.
    
    Majorization with the vectors $\mu$,  not only improves the existing bound on $p$-norms, but also improves lower bounds for other quantities like Rényi entropies. A  general lower bound  for the Rényi entropies can be obtained by weak subadditivity \cite{van2002Renyi},
 \begin{equation*}
     S_{\alpha}(\rho_A)\geq S_{\alpha}(\rho_{AB})-\ln d_B, \qquad S_{\alpha}(\rho_B)\geq S_{\alpha}(\rho_{AB})-\ln d_A \, ,
 \end{equation*}
and using that  Rényi entropies are positive,
 \begin{subequations}
 \begin{align}
      S_{\alpha}(\rho_A)+ S_{\alpha}(\rho_B)& \geq S_{\alpha}(\rho_{AB})-\ln \min\{ d_A,d_B\}\\ \label{eq:lowerboundRenyiDimension}
      &=S_{\alpha}(\rho_{AB})-\ln 3\, . 
 \end{align}
 \end{subequations}
   
    For $\alpha>1$, we can use that $\ln(x)+\ln(y)\leq \ln(x+y)$ for $0\leq x,y\leq 1$ to obtain that $S_{\alpha}(\rho_A)+S_{\alpha}(\rho_B)\geq S_{\alpha}(\rho_A\oplus \rho_B)$. Solving the QP (with the constraint that all eigenvalues must be positive) one can improve the dimension-dependent lower bound \eqref{eq:lowerboundRenyiDimension} for any $\alpha$ greater than 4, as one can see in Figure \ref{fig:AlphaRenyi}. Note that in this case the QP for $\mu_{I}$ is not feasible.
   % \begin{figure}
    %    \centering
     %   \includegraphics[width=0.7\textwidth]{RenyiMIBound.png}
      %  \caption{Lower bounds for subadditivity of Rényi entropies}
      %  \label{fig:AlphaRenyi}
    %\end{figure}
\end{example}

In a similar spirit, in the case $d_1<d_2$, one can also use the sufficient conditions given in e.g. Proposition \ref{prop:SufficientCaseDifferentDimensions} to create a QP which, when feasible, provides bounds for every Schur-convex/concave functional. 

An implementation of the quadratic programs developed in this section is available in \cite{QPs} and on  \href{https://github.com/pashteiner/partial-trace-inequalities}{GitHub}. The code reproduces the computations in Example \ref{example:numerics}, provides additional examples, and allows users to input arbitrary spectra and compute the corresponding bounds obtained by our approach.

\section{Optimal bounds for singular values of both partial traces}\label{sec:SVD}

After thoroughly studying the maxima of Schur-convex functions along unitary orbits for self-adjoint matrices, the next natural step is to also look for optimal bounds as a function of singular values, that is,  considering now a problem of the form
\begin{equation}
    \max_{U,V \in \U{d_1,d_2}}f(\tr_1[UCV]\oplus \tr_2[UCV])\, .
\end{equation}
 In the spirit of the previous sections, we will look for weak majorization of the spectra of the direct sum of partial traces in terms of an associated diagonal matrix with  singular values in the  decreasing order, that is
\begin{equation}\label{eq:ConjectureWeakMajorizationTwoPartialTracesSVD}
  \sigma(\tr_1 [C] \oplus \tr_2 [C])\preceq_{\omega}  \sigma(\tr_1 [\Sigma] \oplus \tr_2 [\Sigma])\, .
\end{equation}

  Consider, for $i=1,2$, unitaries $U_1,V_1 \in \U{d_2}$ and $U_2,V_2 \in \U{d_1}$ such that both $U_2 \tr_2 [C] V_2$ and $U_1\tr_1 [C] V_1$ are the diagonal singular-values matrices of $\tr_2 [C]$ and $\tr_1 [C]$ respectively. Define the matrix $\hat C=(U_2\otimes U_1)C(V_2\otimes V_1)$ satisfying $\sigma(C)=\sigma(\hat{C})$. If we write the singular value decomposition of both partial traces
 \begin{equation}
     \tr_1 [C]=\sum_{i=1}^{d_2}\sigma^{(1)}_i \vert v_i \rangle \langle w_i\vert, \quad   \tr_2 [C]=\sum_{i=1}^{d_1}\sigma^{(2)}_i \vert x_i \rangle \langle y_i\vert \, ,
 \end{equation}
then 
 \begin{align*}
     \sigma^{(1)}_i &=\tr[\tr_1 [C] \vert w_i \rangle \langle v_i \vert]\\ &=\tr[ C (\id_1 \otimes 
 \vert w_i \rangle \langle v_i \vert)]\\ &=\tr[(U_2\otimes U_1)^*\hat{C}(V_2\otimes V_1)^* (\1_1 \otimes \vert w_i  \rangle \langle v_i \vert)]\\ &=\tr [\hat{C}(V_2^*U_2^*\otimes P_i)]\, ,
 \end{align*}
where $P_{i} \in \L{d_2}$ is a matrix of zeros with a $1$ in the $i$-th diagonal entry, and similarly
 \begin{equation*}
     \sigma^{(2)}_j=\tr[\hat{C}(Q_j\otimes V_1^*U_1^*)]
 \end{equation*}
 where $Q_j \in \L{d_1}$ is a matrix of zeros with a $1$ in the $j$-th diagonal entry. Therefore, the expression that we need to consider to show majorization of the singular values is of the form
 \begin{equation}\label{eq:SumSingValues}
     \sum_{i=1}^{k_1}\sigma_i^{(1)}+\sum_{j=1}^{k_2}\sigma_j^{(2)}=\tr\left[\hat{C}\left( V_2^*U_2^*\otimes\sum_{i=1}^{k_1}  P_i+\sum_{j=1}^{k_2}Q_j\otimes V_1^*U_1^* \right) \right]\, ,
 \end{equation}
for $k_1 \leq d_2$ and $k_2\leq d_1$. To be able to bound this term, we show a majorization result in the spirit of \cite[Lemma 1]{rico2025partial}.

 \begin{lemma}\label{lem:SVD}
     Let   $\{W_i\}_{i=1}^{k_1} \subset \U{d_1}$, $\{\widetilde{W}_j\}_{j=1}^{k_2} \subseteq \U{d_2}$ be unitary matrices,  $k_1\leq d_2$ and $k_2 \leq d_1$. If we define
     \begin{equation}
         S=\sum_{i=1}^{k_1} W_i\otimes P_{i}+\sum_{j=1}^{k_2} Q_{j}\otimes \widetilde{W}_j\, , 
     \end{equation}
    then
     \begin{equation*}
         \sigma(S)\preceq_{\omega}(\underbrace{2,\hdots, 2}_{k_1k_2},\underbrace{1,\hdots,1,}_{k_1d_1+k_2d_2-2k_1k_2}0,\hdots ,0)^T=:v_S \, .
     \end{equation*}
 \end{lemma}
\begin{proof}
    This proof consists of a modification of  \cite[Lemma 1]{rico2025partial}. Let $\{f_l\}_l$ be an orthonormal basis of $\mathbb{C}^{k_1+k_2}$, then we can write $S=LR$ with 
    \begin{eqnarray}
        L &:=& \sum_{i=1}^{k_1}  W_i \otimes P_{i}\otimes \langle f_i \vert +\sum_{j=1}^{k_2}Q_j \otimes \widetilde{W}_j \otimes \langle f_{k_1+j} \vert ,\nonumber\\
        R &:=& \sum_{i=1}^{k_1}  \1 \otimes P_i \otimes \vert f_i \rangle +\sum_{j=1}^{k_2}Q_j \otimes \1 \otimes \vert f_{k_1+j} \rangle \nonumber \, .
    \end{eqnarray}
    We apply now the Cauchy-Schwarz inequality \cite[IX.5]{Bhatia.1997} on an arbitrary unitarily invariant norm $\unorm{\cdot}$, and obtain 
    \begin{subequations}
    \begin{align}
        \unorm{S} &= \unorm{LR}\\ &\leq \unorm{L^*L}^{\frac{1}{2}}\unorm{R^*R}^{\frac{1}{2}}\\ &=\unorm{L L^*}^{\frac{1}{2}}\unorm{R^*R}^{\frac{1}{2}}\nonumber\\
        &= \unorm{\sum_{i=1}^{k_1}  \1 \otimes P_{i}+\sum_{j=1}^{k_2}Q_j \otimes \1}
    \end{align}
       \end{subequations}
If, in particular, we consider the Ky Fan norms,  we obtain the weak majorization relation 
\begin{equation*}
\begin{split}
    \sigma(S)&\preceq_{\omega} \sigma\left(\sum_{i=1}^{k_1}  \1 \otimes P_i+\sum_{j=1}^{k_2}Q_j \otimes \1 \right)^T\\
    &=(\underbrace{2,\hdots, 2}_{k_1k_2},\underbrace{1,\hdots,1,}_{k_1d_1+k_2d_2-2k_1k_2}0,\hdots ,0)^T \, ,
\end{split}
\end{equation*}
 where we used Remark \ref{rem:MajorizationPartialTracesByDiagLambda} in the last equality.
\end{proof}

Continuing the previous argument, consider the matrix
\begin{equation}
    S=V_2^*U_2^*\otimes\sum_{i=1}^{k_1}  P_i+\sum_{j=1}^{k_2}Q_j\otimes V_1^*U_1^*\, ,
\end{equation}
that appears in the expression \eqref{eq:SumSingValues}. We can now apply the previous Lemma to $S$ and the von Neumann trace theorem \cite[Theorem 2]{mirsky1959trace}  to obtain
 \begin{subequations}
     \begin{align}
\sum_{i=1}^{k_1}\sigma_i^{(1)}+\sum_{j=1}^{k_2}\sigma_j^{(2)}&= \tr[\hat{C}S]\\
&\leq \sum_{i=1}^{d_1d_2} \sigma(\hat{C})_i\sigma(S)_i  \\
&\leq \sum_{i=1}^{d_1d_2}  \sigma(C)_i (v_S)_i \\
&\leq \sum_{i=1}^{k_1d_1+k_2d_2-k_1k_2}\sigma_i+\sum_{j=1}^{k_1k_2}\sigma_j\,  .
\end{align}
 \end{subequations}
 Once this upper bound for the singular values of partial traces has been obtained, we can argue in a similar way to Propositions \ref{prop:NewSufficientConditionsSquareCase}, \ref{prop:sufficient_conditions_small_dimensions}, \ref{prop:SufficientCaseDifferentDimensions}  and \ref{prop:sufficientRank3}  to obtain sufficient conditions for majorization with singular values, because they follow from \eqref{eq:estimated_by_two_sums_of_eigenvalues}. By replacing $\lambda$ with $\sigma$ in the respective proofs we obtain the following results.
 \begin{corollary}\label{coro:SufficientConditionsSquareCaseSVD}
    Let $d \geq 2$ and $C \in  \L{d, d}$  satisfying $\sigma_n=\hdots=\sigma_{d^2-4+n}$ for some $n \in \{1,2,3,4\}$. Then,
    \begin{equation*}
        \sigma(\tr_1 [C] \oplus \tr_2 [C])\preceq_{\omega}  \sigma(\tr_1 [\Sigma] \oplus \tr_2 [\Sigma]) \, .
    \end{equation*}   
 \end{corollary}
\begin{corollary}\label{coro:SufficientConditionsRank3SVD}
       Let $d_1,d_2 \geq 2$ and $C \in  \L{d_1, d_2}$ satisfying one of the following conditions:
       \begin{enumerate}
           \item $\rk(C)\leq 3$\, .
           \item $\sigma_2=\hdots=\sigma_{d_1d_2-1}$\, .
       \end{enumerate}
        Then,
    \begin{equation*}
        \sigma(\tr_1 [C] \oplus \tr_2 [C])\preceq_{\omega}  \sigma(\tr_1 [\Sigma] \oplus \tr_2 [\Sigma]) \, .
    \end{equation*}   
 \end{corollary}

 Under any of the assumptions of the previous Corollaries, for every monotonically increasing Schur-convex function $f$ such that $\tr_1 [C] \oplus \tr_2 [C], \tr_1 [\Sigma] \oplus \tr_2 [\Sigma] \in \text{dom}(f)$, we conclude that
 \begin{equation*}
     f(\tr_1 [C] \oplus \tr_2 [C])\leq f(\tr_1 [\Sigma] \oplus \tr_2 [\Sigma])\, .
 \end{equation*}

 \begin{rem}[Optimal bounds for unitarily invariant norms]
     Under any of the assumptions of Corollaries \ref{coro:SufficientConditionsSquareCaseSVD} and \ref{coro:SufficientConditionsRank3SVD}, we obtain that for any unitarily invariant norm $\unorm{\hspace{2pt}\cdot \hspace{2pt}}$
     \begin{eqnarray}
         \unorm{\tr_1 [C] \oplus \tr_2 [C]}\leq  \unorm{\tr_1 [\Sigma] \oplus \tr_2 [\Sigma]}\, ,
     \end{eqnarray}
     due to Fan's dominance theorem \cite[Theorem IV 2.2]{Bhatia.1997}.
     In the particular case of the Schatten 
     $p$-norms $\Vert \hspace{2pt}\cdot \hspace{2pt} \Vert_p$, one can obtain the inequality
     \begin{equation}\label{ineq:2DistillabilityP}
         \Vert \tr_1 [C] \Vert_p^p+\Vert \tr_2 [C] \Vert_p^p \leq \Vert \tr_1 [\Sigma] \Vert_p^p+\Vert \tr_2 [\Sigma] \Vert_p^p \, ,
     \end{equation}
     for any $p \geq 1$.
     
     For matrices $C$ with rank $r\leq 3$, the bound \eqref{ineq:2DistillabilityP} coincides with the one  obtained in \cite[Prop. 9]{rico2025partial}  
     after the application of the monotone Schur-convex function $x^p$, $p\geq 1$:
     \begin{equation}
         \Vert \tr_1 [C]\Vert_p^p+\Vert \tr_2 [C]\Vert_p^p\leq \Vert C\Vert_p^p+\Vert C \Vert_1^p\, .
     \end{equation}
     This shows an alternative way to obtain  \cite[Cor. 3]{rico2025partial} which concerns the distillability of Werner states.
     
     However, in some cases like e.g. $p=d_1=d_2=2$ and $C$ with singular values $\sigma_1> \hdots> \sigma_4>0$, \eqref{ineq:2DistillabilityP} presents an improvement:
     \begin{equation}
         \Vert \tr_1 [\Sigma] \Vert_2^2+\Vert \tr_2 [\Sigma] \Vert_2^2=2\Vert C \Vert_2^2+2(\sigma_1+\sigma_4)(\sigma_2+\sigma_3)<\Vert C \Vert_2^2+\Vert C \Vert_1^2
     \end{equation}
 \end{rem}

 \begin{rem}
     In an analogous way as we saw in Section \ref{sec:QP} we can use the   QP  \eqref{eq:SDPSameDimensions} to obtain general upper bounds for Schur-convex functions. In this case, we have to add the extra restriction that every component must be positive and we have to replace the trace preservation condition with the lower bound by the one-norm, i.e. $b_{eq}^Tx \geq \Vert C \Vert_1$.
 \end{rem}

\section{Necessary rank conditions for majorization with diagonal matrix}\label{sec:BothPartialTracesSameDimension}

Although there are certain cases where the maximum of the problem \eqref{eq:ProblemMaximzerBothPartialTraces} is attained at a diagonal matrix, in general this is not true. In this section, we will discuss when it is impossible to find such majorization by diagonal in terms of the rank of the matrix.

To construct  such counterexamples, we are going to consider orthogonal transformations applied to diagonal matrices in a structured manner: each transformation acts non-trivially only on a $ 2 \times 2 $ principal submatrix. More precisely, given a (not necessarily decreasing) spectrum $\lambda' \in \RRR^n$, indices $i, j \leq n$ and $0 < \alpha < |\lambda'_i - \lambda'_j|$, we define an orthogonal transformation of the form 
\begin{equation*} U_{i,j}(\alpha) = I_n + (e_i \hspace{5pt} e_j)(V_{i,j}(\alpha) - I_2)(e_i \hspace{5pt} e_j)^T  \, ,
\end{equation*}
where $(e_i \hspace{5pt} e_j)$ is the $n\times 2$ matrix formed by the $i$-th and the $j$-th standard basis vectors of $\mathbb{R}^n$ and  $ V_{i,j}(\alpha) \in \L{2}$ is an orthogonal matrix satisfying
\begin{equation*}
     V_{i,j}(\alpha)\begin{pmatrix}
         \lambda'_i & 0\\
         0 & \lambda'_j
     \end{pmatrix} V_{i,j}(\alpha)^T=\begin{pmatrix}
         \lambda'_i-\sign(\lambda'_i - \lambda'_j) \alpha & *\\
           * & \lambda'_j+\sign(\lambda'_i - \lambda'_j) \alpha
     \end{pmatrix}.
\end{equation*}
The existence of such $V_{i, j}(\alpha)$ follows from Horn's theorem \cite[Theorem II.9.B.2]{Marshall.2011} that states that for every $v, \lambda' \in \RR^n$ with $v \preceq \lambda'$ there exists a real symmetric matrix $S$ with diagonal $v$ and spectrum $\lambda'$.  

From now on, given a not necessarily decreasing spectrum $\lambda' \in \RR^{d_1d_2}$ with associated diagonal matrix $\Lambda'=\diag(\lambda')$, we introduce the following operation
\begin{equation}\label{eq_T_alpha}
    T_{\alpha, i, j}(\Lambda') \coloneqq U_{i,j}(\alpha) \Lambda' U_{i,j}(\alpha)^T \in \L{d_1, d_2}.
\end{equation}
This operation preserves the spectrum  and the diagonal entries of $T_{\alpha, i, j}(\Lambda')$ are obtained from $\lambda'$ via a $T$-transform of $\alpha$ between $\lambda'_{i}$ and $\lambda'_j$ (see e.g. \cite[Section 2.B]{Marshall.2011}). The only two off-diagonal entries of $T_{\alpha, i, j}(\Lambda')$ are located in positions $ij$ and $ji$ and their exact value  will be irrelevant for this work, if these entries do not play any role after taking partial trace. This occurs if both partial traces are still diagonal:
\begin{equation}\label{eq:partial_traces_T_alpha_ diagonal}
    \begin{array}{l}
        \tr_1 [T_{\alpha, i, j}(\Lambda')] \text{ is diagonal, whenever } \left\lceil i / d_2\right\rceil \neq \left\lceil j / d_2\right\rceil;\\
        \tr_2[ T_{\alpha, i, j}(\Lambda')] \text{ is diagonal, whenever } i \not\equiv j \mod d_2.
    \end{array}
\end{equation}

\bigskip

However,  reasoning in terms of the spectrum can be sometimes a bit cumbersome. For that reason, in some specific cases, a more general approach can be used to construct such counterexamples. Given a real spectrum $\lambda$, take arbitrary vector $c \preceq \lambda$. By Horn's theorem, there exists a matrix $C$ with diagonal $c$ and spectrum $\lambda$. While we cannot easily say what is the spectrum of $\tr_i [C]$, the diagonals are obtained by \eqref{eq:diag:one,two}. Since the diagonal of a matrix is always majorized by its spectrum, the necessary conditions can be generalized from spectra to diagonals. The word generalized is justified: by Lemma \ref{lem:C_hat} every spectrum is a diagonal in an appropriate basis.

\begin{proposition}[Impossible ranks]\label{prop:ImpossibleSquareGeneralRanks}
    Let $C \in \L{d,d}$, $d\geq 3$, be a positive semidefinite matrix with rank $r\geq 4$ and spectrum $\lambda(C)$. If  one of the following conditions holds:
    \begin{itemize}
        \item[(i)] $r<d^2$ is a perfect square.
        \item[(ii)] $r< (d-\sqrt{d})^2$, $0\neq r-\lfloor \sqrt{r} \rfloor^2\leq \lfloor \sqrt{r}\rfloor$.
        \item[(iii)] $r< (d-\sqrt{2d})^2$ and $r-\lfloor \sqrt{r} \rfloor^2> \lfloor \sqrt{r}\rfloor$,
    \end{itemize}
    then there exists a unitary matrix $U\in \U{d,d}$ such that 
    \begin{equation*}\lambda( \tr_1[U CU^*]\oplus \tr_2[UCU^*])\npreceq \lambda(\tr_1[\Lambda_{\pi}]\oplus \tr_2[\Lambda_{\pi}] ) \text{ for every } \pi \in S_{d^2}\, ,\end{equation*}
    where $S_{n}$ denotes the symmetric group of degree $n$.
\end{proposition}
\begin{proof}
     Let $r$ be the rank of $C$. For a permutation $\pi$ we will denote by $\Lambda_\pi$ the diagonal matrix with $(\Lambda_{\pi})_{ii}=\lambda_{\pi(i)}$.  The first step of the proof consists of characterizing the diagonal matrices $\Lambda_{\pi}$ such that the number of zero entries of $\lambdasum{\Lambda_{\pi}}$ is maximal. Given a permutation $\pi \in S_{d^2}$ define
    \begin{equation*}
        z_{\pi}^{(1)}= \text{number of non-zero elements of } \lambda^{(1)}(\Lambda_{\pi})\, ,
    \end{equation*}
        \begin{equation*}
        z_{\pi}^{(2)}= \text{number of non-zero elements of } \lambda^{(2)}(\Lambda_{\pi})\, ,
    \end{equation*}
  which satisfy $1 \leq  z_{\pi}^{(1)}, z_{\pi}^{(2)}\leq d \, ,  \text{ and } z_{\pi}^{(1)}z_{\pi}^{(2)}\geq r$. Observe that, according to  \eqref{eq:computationPartialTrace}, $z_{\pi}^{(1)}$ corresponds to the number of non-zero diagonal elements of the sum of the diagonal blocks, and $z_{\pi}^{(2)}$ corresponds to the number of non-zero diagonal blocks of $\Lambda_{\pi}$. The maximum number of zeros for $\lambdasum{\Lambda_{\pi}}$ can be found by computing
    \begin{equation}
        2d-\min_{\pi \in S_{d^2}}(z_{\pi}^{(1)}+z_{\pi}^{(2)}) \, .
    \end{equation}
 
    If $r$ is a perfect square, then the optimal solution is clearly achieved whenever $z_{\pi}^{(1)}=z_{\pi}^{(2)}=\sqrt{r} \eqqcolon u$. When $r$ is not a perfect square, let $u \in \mathbb{N}$ such that  
    $$
    u^2<r<(u+1)^2\, .
    $$
    In this case,
    \begin{equation}
        \text{maximum number of zeros of }\lambdasum{\Lambda_{\pi}}=\left\{\begin{array}{ccc}
             2d-2u-1& \text{if} & r-u^2\leq u  \\
             2d-2u-2& \text{if} & r-u^2>u
        \end{array}\right.
    \end{equation}
    Since by assumptions, $r\geq 4$, it must be that $u \geq 2$, which we will use in the following. We  study each of the three cases in the statement separately. 

    For the case (i), suppose that $r$ is a perfect square. Then $u=\sqrt{r} < d$. %=z_{\pi}^{(1)}=z_{\pi}^{(2)}$As a consequence $\Lambda_{\pi}$\footnote{Note that in this case $\Lambda_\pi$ does not denote the diagonal matrix obtained from $\Lambda$ by permutation $\pi$ in each entry. \PCRcom{Remove}\PScom{ We need to deal with this problem}} must have diagonal
    Consider a diagonal matrix $D$ in the unitary orbit of $C$, defined by \begin{equation}
        \diag(D)^T=\left( \bigoplus_{i=1}^u (\underbrace{\lambda_{(i-1)u+1}, \hdots, \lambda_{iu}}_u,0, \hdots,0)\right)\bigoplus\left( \bigoplus_{i=u+1}^d(0, \hdots, 0)\right) \, .
    \end{equation}
    
     Consider for $0<\alpha< \lambda_r$
    \begin{equation*}
        C_{\alpha}=T_{\alpha,u+1,d+1}(D)\, ,
    \end{equation*}
    The number of zero elements of $\lambdasum{C_{\alpha}}$ is $2d-2u-1$. Since $\alpha<\lambda_r$, majorization $\lambdasum{C_{\alpha}} \preceq \lambdasum{\Lambda_{\pi}}$, for some $\pi \in S_{d^2}$, requires $\lambdasum{\Lambda_{\pi}}$  to have $2d-2u$ zeros. This forces $z_{\pi}^{(1)}=z_{\pi}^{(2)} = u$ and therefore  %$\Lambda_{\sigma}$ to be of the form \begin{equation}\label{eq:necessaryRank1} \diag(\Lambda_{\sigma})^T=\left( \bigoplus_{i=1}^u (\underbrace{\lambda_{\pi((i-1)u+1)}, \hdots, \lambda_{\pi(iu)}}_u,0, \hdots,0)\right)\bigoplus\left( \bigoplus_{i=u+1}^d(0, \hdots, 0)\right) \,
    % \end{equation}
    % for some $\pi\in S_{r}$ and
    % up to permutations of diagonal blocks and jointly shifting the eigenvalues in each block, which leave invariant $\lambdasum{\Lambda_{\sigma}}$. However, in this case,
\begin{equation}
    \max \lambdasum{C_{\alpha}}=\lambda_1+\hdots+\lambda_u+\alpha> \lambda_1+\hdots+\lambda_u\geq\max \lambdasum{\Lambda_{\pi}}\, ,
\end{equation}
which is a contradiction.

If item (ii) holds, then $r-u^2\leq u$ and the maximum number of zeros that $\lambdasum{\Lambda_{\pi}}$ can achieve is $2d-2u-1$. Assume without loss of generality that $z_{\pi}^{(1)}\leq z_{\pi}^{(2)}$. Then the possible configurations that provide $2d-2u-1$ zeros are $(z_{\pi}^{(1)},z_{\pi}^{(2)})=(u,u+1),(u-1,u+2), \hdots,(u-k,u+k+1)$ as long as $r\leq (u-k)(u+k+1)$ and $u+k+1\leq d$. Let
\begin{equation*}
    k_{\max}=\max\{k \in \mathbb{N}\cup \{0\}: r\leq (u-k)(u+k+1), u+k+1\leq d\}\, ,
\end{equation*}
and assume that $u+k_{\max}+1<d$ (below we will show that this follows from (ii)). Consider a diagonal matrix $D$ in the unitary orbit of $C$, defined by
\begin{equation}
\begin{split}
    \diag(D)^T= &\bigoplus_{i=1}^{u-k_{\max}-1}\left(\underbrace{\lambda_{(i-1)(u+k_{\max}+1)+1},\hdots,\lambda_{i(u+k_{\max}+1)}}_{u+k_{\max}+1},\underbrace{0,\hdots,0}_{d-u-k_{\max}-1} \right)\\
    &\hspace{20pt}\bigoplus \left(\underbrace{\lambda_{(u+k_{\max}+1)(u-k_{\max}-1)+1},\hdots, \lambda_r}_{r-(u+k_{\max}+1)(u-k_{\max}-1)}, 0, \hdots,0 \right)\\
    &\hspace{5pt}\bigoplus_{i=u-k_{\max}+1}^d\left(\underbrace{0,\hdots,0}_d\right)
\end{split}
\end{equation}
and consider, for $0<\alpha<\lambda_r$ the matrix
\begin{equation*}
    C_{\alpha}=T_{\alpha,u+k_{\max}+2,d+1}(D) \, ,
\end{equation*}
As before, the number of zero elements of $\lambdasum{C_{\alpha}}$ is $2d-2u-2$ and a non-zero element $\alpha \leq \lambda_r$. Therefore we will need to majorize with $\lambdasum{\Lambda_{\pi}}$ with $2d-2u-1$ zeros, for some $\pi \in S_{d^2}$. This is however not possible since 
\begin{equation}
    \max \lambdasum{C_{\alpha}}=\lambda_1+\hdots+\lambda_{u+k_{\max}+1}+\alpha> \max \lambdasum{\Lambda_{\pi}}\, ,
\end{equation}
for every $\Lambda_{\pi}$ with $2d-2u-1$ zeros, because  every entry of $\lambdasum{\Lambda_\pi}$ contains at most $u+k_{\max}+1$ summands.

All that is left to prove is that (ii) implies $u+k_{\max}+1<d$.  From our assumption $r\leq (u-k_{\max})(u+k_{\max}+1)$, we obtain that $k_{\max}(k_{\max}+1)\leq u(u+1)-r$. We want to show  that $k_{\max} <d-u-1$ which is equivalent to show that $k_{\max}(k_{\max}+1)<(d-u-1)(d-u)$, since the function $f(x)=x(x+1)$ is strictly monotone  increasing for $x\geq 0$. Therefore, it is sufficient to prove that
\begin{equation*}
    u(u+1)-r<(d-u)(d-u-1)\, ,
\end{equation*}
which is equivalent to
\begin{equation*}
    r>d(1+2u-d)\, .
\end{equation*}
Since $u\leq \sqrt{r}$, it suffices to show
\begin{equation*}
    r>d(1+2\sqrt{r}-d)\, .
\end{equation*}
This is a second order inequality in terms of $\sqrt{r}$ which holds if and only if $\sqrt{r}<d-\sqrt{d}$ or $\sqrt{r}>d+\sqrt{d}$. From the first constraint we deduce that if $r<(d-\sqrt{d})^2$, then $u+k_{\max}+1<d$.

Finally, if item (iii) holds, then $r-u^2>u$
 and we can achieve at most $2d-2u-2$ zeros. The possible configurations in this case are $(z_{\pi}^{(1)},z_{\pi}^{(2)})=(u+1,u+1),(u,u+2),\hdots,(u-k,u+2+k)$, for $k \geq 0$ as long as $r\leq (u-k)(u+k+2)$ and $u+k+2\leq d$. Following the same steps as in (ii), we construct the matrix $C_{\alpha}$ in an analogous way such that $\lambdasum{C_{\alpha}}$ is not majorized by any vector associated to a diagonal matrix, whenever $u+k_{\max}+2<d$. Repeating the proof of  item (ii), it suffices to find the ranks $r$ such that
 \begin{equation*}
     r>d(2+2\sqrt{r}-d)\, ,
 \end{equation*}
 and this occurs for $r<(d-\sqrt{2d})^2$.
\end{proof}

    Since $f(x)\leq f(y)$ for every Schur-convex function implies $x \preceq y$ \cite[Chapter 4.B]{Marshall.2011}, whenever a matrix $C$ satisfies any of the conditions stated in Proposition \ref{prop:ImpossibleSquareGeneralRanks}, then there exists a  Schur-convex function $f$ such that the problem \eqref{eq:ProblemMaximzerBothPartialTraces} is not attained at a diagonal matrix. Nevertheless, for certain specific Schur-convex functions, the optimum may still be achieved at a diagonal element of the orbit, as occurs for the von Neumann entropy (see Remark \ref{rem:LowerBoundMI}).

    A case that Proposition \ref{prop:ImpossibleSquareGeneralRanks} always excludes is $4\leq r \leq d$ and as  a consequence we obtain the following corollary.

\begin{corollary}\label{coro:SmallRanksSquareCase}
    Let $C \in \L{d,d}$ be positive semidefinite with $d\geq 4$ and with rank $4\leq r \leq d$. Then, there exists a unitary matrix $U\in \U{d,d}$ such that  
   \begin{equation*}\lambdatrsum{[U CU^*]}\npreceq \lambdatrsum{[\Lambda_{\pi}]} \text{ for every } \pi \in S_{d_1d_2}.\end{equation*}
\end{corollary}

 We will show later in \cref{lema:r<=d_2} that this statement also holds true for different dimensions of the subsystems.    
Having established a result that rules out a broad range of ranks, it is natural to ask for which ranks the solution of \eqref{eq:ProblemMaximzerBothPartialTraces} can be attained in a diagonal matrix for every Schur-convex functional. Corollary \ref{coro:SufficientConditionsd<4SquareCase} shows that this is indeed always the case for $r\leq 3$. The next result shows that, for most ranks, it is not possible to guarantee that the optimizer of \eqref{eq:ProblemMaximzerBothPartialTraces} is attained for an arbitrary spectrum of such fixed rank. 

\begin{corollary}[Characterization for equal non-zero spectrum in large dimensions]\label{coro:d>24}
    Let $C \in \L{d,d}$, $d \geq 24$,  be a positive semidefinite matrix with rank $r$ and $\lambda(C)=(\lambda, \hdots, \lambda,0, \hdots, 0)$. There exists a permutation $\pi \in S_{d^2}$ such that $\lambda^{(1,2)}(UCU^*) \preceq \lambda^{(1,2)}(\Lambda_{\pi})$ for every $U\in \U{d,d}$, if and only if the rank  satisfies one of the following conditions:
    \begin{itemize}
        \item[(i)] $r \leq 3$.
        \item[(ii)] $r \geq d^2-3$.
    \end{itemize}
\end{corollary}
\begin{proof}
     Sufficiency is provided in \cref{prop:NewSufficientConditionsSquareCase} and \cref{coro:SufficientConditionsd<4SquareCase}. For the necessary conditions, let $C$ be a matrix with rank $r$ and non-zero eigenvalues equal to $\lambda$ and suppose that  that there exists a unitary $U\in \U{d,d}$ such that for every permutation $\pi \in S_{d^2}$,  \begin{equation}\label{eq:d>=24:no_maj}\lambda^{(1,2)}(UCU^*) \npreceq \lambda^{(1,2)}(\Lambda_{\pi}).\end{equation} We show first that if for some rank $r$ majorization by diagonal does not hold, it does not hold for $d^2-r$ either.
     
     Since both local dimensions are equal, relation \eqref{eq:d>=24:no_maj} is equivalent to
    \begin{equation*}
        \lambda^{(1,2)}\left( U(C-\lambda I)U^*  \right)\npreceq \lambda^{(1,2)}\left( \Lambda_{\pi}-\lambda I \right) \, ,
    \end{equation*}
    since it is obtained from \eqref{eq:d>=24:no_maj} by adding the vector $-\lambda d e$, where $e$ is the vector of ones on both sides of the relation. Since multiplying by minus one preserves majorization, we obtain 
    \begin{equation*}
        \lambda^{(1,2)}\left( U(\lambda I-C)U^*  \right)\npreceq \lambda^{(1,2)}\left( \lambda I-\Lambda_{\pi} \right) \, .
    \end{equation*}
    However, since the operation $C \mapsto \lambda I- C$ is bijective and maps $C$ of rank $r$ to a rank $d^2-r$ matrix,  we conclude that majorization by diagonal does not follow for matrices with rank $d^2-r$ with non-zero equal spectrum. Therefore we can assume that $r \leq \frac{d^2}{2}$.

    To conclude, notice that for $d\geq 24$, $(d-\sqrt{2d})^2\geq \frac{d^2}{2}$, so by \cref{prop:ImpossibleSquareGeneralRanks} the result follows.
\end{proof}

The previous result shows that for $d\geq 24$ there are strict constraints on the rank for an optimal solution valid for any spectrum with such rank. However, a similar  phenomenon occurs for $3\leq d<24$. Again, we need to study which ranks are also not possible for the case where all the eigenvalues are equal, apart from those given in \cref{prop:ImpossibleSquareGeneralRanks}. For this purpose,  we will need the following preliminary result.

\begin{lemma}\label{lem:rankBetweenMultipliesd_2} Let $2\leq d_1 \leq d_2$ and let $C\in \L{d_1, d_2}$ be positive semidefinite. If $(k-1)d_2 < \rk(C)\leq k(d_2 - 1)$ for some $2 \leq k \leq d_1$, then there exists a unitary $U \in \U{d_1, d_2}$ such that \[\lambdatrsum{[UCU^*]}\npreceq \lambdatrsum{[\Lambda_{\pi}]} \text{ for every } \pi \in S_{d_1d_2}.\]
\end{lemma}
\begin{proof}
See Lemma \ref{lemAPP:rankBetweenMultipliesd_2} in the appendix.
\end{proof}

\begin{proposition}[Necessary ranks in small dimensions]\label{prop:d<24}
    Let $C \in \L{d,d}$, $3\leq d<24$, be a positive semidefinite matrix with rank $r$ and $\lambda(C)=(\lambda, \hdots, \lambda,0, \hdots, 0)$. If for every unitary $U \in \U{d,d}$, there exists permutation $\pi \in S_{d^2}$ such that $\lambda^{(1,2)}(UCU^*) \preceq \lambda^{(1,2)}(\Lambda_{\pi})$ holds, then the rank must satisfy one of the following conditions:
    \begin{itemize}
        \item[(i)] $r \leq 3$.
        \item[(ii)] $r \geq d^2-3$.
        \item[(iii)] $r=td$ for some $2 \leq t \leq d-2$.
    \end{itemize}
\end{proposition}
\begin{proof}
   
    Consider  the sets
    \begin{equation*}
        A_d=\{ r \in \{1,\hdots, d^2\} : (k-1)d <r \leq k(d-1) \text{ for some  } 2\leq k\leq d-1 \}\, ,
    \end{equation*}
    and 
    \begin{equation*}
        B_d=\{d^2-r : r \in A_d\}\, .
    \end{equation*}
By \cref{lem:rankBetweenMultipliesd_2}, we know that  the matrices with  ranks in $A_d$ do not satisfy the majorization by diagonal property, and by the discussion  given in the proof of \cref{coro:d>24} neither do the  matrices with ranks in $B_d$. We show that  
\begin{equation}\label{eq:AdUnionBd}
    A_d\cup B_d=\{ r \in \{d,\hdots, d^2-d\} : r\neq td, \text{for some } 1 \leq t \leq d-1\}.
\end{equation}
    The inclusion $"\subseteq "$ is clear. To show  $"\supseteq "$, let $r \in \{d,\hdots, d^2-d\}$ such that $d \nmid r$, and write $r=dq+k$, with $1\leq k,q \leq d-1$. Since
    \begin{equation}\label{eq:ineqResidue}
    qd<r< (q+1)d\, ,
    \end{equation} $r \in A_d$ if and only if $r \leq (q+1)(d-1)$, or equivalently $k\leq  d-1-q$. We need to show then that if $k> d-q-1$, i.e. $k \geq d-q$, then $d^2-r \in A_d$.
    
    Now, condition \eqref{eq:ineqResidue} is equivalent to
    \begin{equation}
        (d-q-1)d<d^2-r < (d-q)d\, ,
    \end{equation}
    so $d^2-r \in A_d$ if and only if $d^2-r\leq (d-q)(d-1)$. This last condition is equivalent to $k \geq d-q$, which shows \eqref{eq:AdUnionBd}.

     Finally, Corollary \ref{coro:SmallRanksSquareCase} and the prior arguments exclude the cases $r\in [4,d]\cup [d^2-d,d^2-4]$.
\end{proof}

In general, for  $3 \leq d <24$ and for a positive semidefinite matrix $C$ with rank $r$ and  spectrum $\lambda(C)=(\lambda, \hdots,\lambda,0,\hdots,0)$, Condition  (iii) of \cref{prop:d<24} can  actually be  sufficient.  For example, if $d=4$ and $r=8$, it can be checked that $\lambda^{(1,2)}(C)\preceq \lambda^{(1,2)}(\Lambda)$. This in particular shows that the characterization of  \cref{coro:d>24} does not necessarily hold for $d < 24$.

Apart from Lemma \ref{lem:rankBetweenMultipliesd_2} we also show in the next result  that for small low ranks larger than three, the majorization of the joint spectra of the partial traces in general dimensions is not attained in any diagonal representative.
\begin{lemma}[Low rank majorization]\label{lema:r<=d_2}
    Let $2\leq d_1 < d_2$,  and let $C \in \L{d_1, d_2}$ be  a positive semidefinite  matrix with $4\leq \rk(C)\leq d_2$. Then there exists a unitary $U\in \U{d_1, d_2}$ such that \begin{equation*}\lambdatrsum{[U CU^*]}\npreceq \lambdatrsum{[\Lambda_{\pi}]} \text{ for every } \pi \in S_{d_1d_2}.\end{equation*}
\end{lemma}
\begin{proof}
    See Lemma \ref{lemaAPP:r<=d_2} in the appendix.
\end{proof}

\section{Bounds for direct sum of  partial traces in n-partite systems of qubits}\label{sec:n-partite}

In this section, our goal is to extend the optimization problem \eqref{eq:ProblemMaximzerBothPartialTraces} to more than two subsystems. To make computations as easy as possible, we will only discuss systems of qubits, that is, we will be interested in finding
\begin{equation}\label{eq:OptimizationProblemNPartite}
    \max_{U \in \operatorname{U}((\mathbb{C}^2)^{\otimes n})}f\left(\bigoplus_{j=1}^n\tr_{\{1,\hdots,n\}\setminus \{j\}} [UCU^*] \right)\, .
\end{equation}

In \cite[Theorem 4.2.3]{klyachko2004quantum}, Klyachko showed also the quantum marginal problem for systems of $n$-qubits, converting the previous optimization problem  over a unitary orbit to an optimization problem of the function $f$ over a polytope. While for a low number of copies this is still manageable (see e.g. \cite[Example 4.2.3]{klyachko2004quantum}), this becomes quickly intractable when $n$ becomes larger. This novel treatment avoid such optimization over the convex polytope by directly showing the maximum for the problem \eqref{eq:OptimizationProblemNPartite} under some assumption on the spectrum.

Denote by $[n]=\{1,\hdots,n\}$. We begin by computing $\tr_{[n] \setminus \{j\}} [\Lambda] \in \L{2}$ for the diagonal matrix $\Lambda$ with eigenvalues in the decreasing order, $1 \leq j \leq n$. Let $\{h_1, \hdots, h_{n-1}\}=[n] \setminus \{j\}$, which satisfy $h_1 <  \hdots <  h_{n-1}$. The action of the partial trace in one system, e.g. on $h_1$ yields the following diagonal  entries
    \begin{equation*}
       \left( \tr_{h_1} [\Lambda] \right)_{i,i}=\lambda_i+\lambda_{i+2^{n-h_1}}.
    \end{equation*}
    By a recursive argument,  if  we denote
    \begin{equation*}
        H=\{b= (b_1,\hdots, b_{n-1}): b_i\in \{ 0,1\} \text{ for every } 1\leq i \leq n-1\},
    \end{equation*}
    then the first eigenvalue can be written as 
\begin{equation}
        \Lambda_{[n]\setminus \{j\}}^{(1)}=\tr_{h_{n-1}}\left( \tr_{h_{n-2}}\left(  \hdots \tr_{h_1} [\Lambda]\right)\right)_{1,1}=\sum_{b \in H}\lambda_{1+b_1 2^{n-h_1}+\hdots +b_{n-1} 2^{n-h_{n-1}}}.
    \end{equation}
    Since the map $b \to x \in \{0,1\}^n$, $x_{h_i}=b_i$, $x_j=0$ is a bijection, we can rewrite
    \begin{equation}\label{eq:partialtracesbinaryexpression}
         \Lambda_{[n]\setminus \{j\}}^{(1)}=\sum_{\substack{x \in \{0,1\}^n\\ x_j=0}} \lambda_x \, , \quad \lambda_x:=\lambda_{1+\sum_k x_k 2^{n-k}}
    \end{equation}
    The second eigenvalue can be computed using the trace and the first eigenvalue:
    \begin{equation*}
        \Lambda_{[n]\setminus \{j\}}^{(2)}=\tr[\Lambda]-\Lambda_{[n]\setminus \{j\}}^{(1)}\, .
    \end{equation*}

\begin{theorem}
    Let $C \in L((\mathbb{C}^2)^{ \otimes n} )$ be a self-adjoint   matrix with eigenvalues $\lambda(C)=(\lambda_1,\hdots \lambda_{2^n})$ in the decreasing order and $[n]=\{1, \hdots,n\}$. If $\lambda_4=\hdots =\lambda_{2^n-3}$, then
    \begin{equation}\label{eq:theo-n-partite}
        \lambda\left(\bigoplus_{j=1}^n \tr_{[n] \setminus \{j\}} [C]\right)\preceq\lambda\left(\bigoplus_{j=1}^n \tr_{[n] \setminus \{j\}} [\Lambda]\right)
    \end{equation}
\end{theorem}
\begin{proof}
   We need  to check that 
    \begin{equation}\label{eq:theo-n-partite2}
        \sum_{i=1}^m \lambda_i^{\downarrow}\left(\bigoplus_{j=1}^n \tr_{[n] \setminus \{j\}} C\right)\leq \sum_{i=1}^m \lambda_i^{\downarrow}\left(\bigoplus_{j=1}^n \tr_{[n] \setminus \{j\}} \Lambda \right)\, ,
    \end{equation}
    for $1\leq m \leq 2n $.

For $n=2$ this is shown in Proposition \ref{prop:sufficient_conditions_small_dimensions} Let $n\geq 3$, $N=2^n$ and  $\lambda:=\lambda_4=\hdots=\lambda_{N-3}$. Denote $\Tilde{C}=C-\lambda I_N$ with $\tr_{[n]\setminus\{j\} }[\Tilde{C}]=\tr_{[n]\setminus\{j\} }[C]-2^{n-1}\lambda I_2$ since we trace out $n-1$ qubits. Hence,
\begin{equation}
    \bigoplus_{j=1}^n \tr_{[n]\setminus \{j\}}[\Tilde{C}]=\bigoplus_{j=1}^n \tr_{[n]\setminus \{j\}}[C]-2^{n-1}\lambda I_{2n}
\end{equation}
Since adding the same scalar multiple of the all-ones vector on both sides of the  majorization relation does not change majorization, we can assume without loss of generality that  $\lambda_4=\hdots=\lambda_{N-3}=0$, so $\lambda_1\geq \lambda_2\geq \lambda_3\geq 0\geq \lambda_{N-2}\geq \lambda_{N-1}\geq \lambda_N$.

Note that we can extend the Local Diagonalization Lemma  to $n$-partite systems: if we let $U_j$ be a unitary matrix diagonalizing $\tr_{[n]\setminus \{j\} }[C]$ and define $\hat{C}=(U_1\otimes \hdots \otimes U_n)C(U_1\otimes \hdots \otimes U_n)^*$, then
\begin{enumerate}
    \item [1.] $\lambda(C)=\lambda(\hat{C})$.
    \item[2.] $\lambda(\tr_{[n]\setminus \{j\}} [C])=\lambda(\tr_{[n]\setminus \{j\}} [\hat{C}])$, for every $1\leq j \leq n$.
    \item[3.] The matrix $\tr_{[n]\setminus \{j\}} [\hat{C}]$ is diagonal for every  $1\leq j \leq n$.
\end{enumerate}
Denote by $\lambda^{(1)}_j \geq \frac{1}{2}\tr[C]\geq \lambda^{(2)}_j$  the two eigenvalues of $\tr_{[n]\setminus \{j\}} [\hat{C}]$  for every  $1\leq j \leq n$. Then, making use of the notation in \eqref{eq:partialtracesbinaryexpression}, we can write the sum of the $m$ largest eigenvalues of the direct sum of partial traces, $m\leq n$ as  
\begin{equation}
    \sum_{\substack{ j\in J} }\lambda_j^{(1)}=\sum_{\substack{ j\in J  }}\sum_{\substack{x \in \{0,1\}^n\\ x_j=0}} \hat{c}_x.
\end{equation}
for some  $J \subseteq [n]$ with $\vert J\vert =m$. Let $S_{k,J}$ be the level set $S_{k,J}=\{ x \in \{0,1\}^n: \sum_{j\in J} x_j=k\}$. Then,
\begin{subequations}
\begin{align}
    \sum_{j \in J}\sum_{\substack{x \in \{0,1\}^n\\ x_j=0}} \hat{c}_x &=\sum_{\substack{x \in \{0,1\}^n}}\left(\sum_{j \in J} \1_{\{x_j=0\}}\right)\hat{c}_x\\
    &=\sum_{\substack{x \in \{0,1\}^n}}\left(\sum_{j \in J} 1-x_j \right)\hat{c}_x\\
    &=\sum_{\substack{x \in \{0,1\}^n}}\Bigg( \underbrace{m-\sum_{j \in J}x_j}_{\omega_J(x)} \Bigg) \hat{c}_x
\end{align}
\end{subequations}

For $r=0,\hdots,m$ the value $m-r$ occurs $2^{n-m}\binom{m}{r}$ times, i.e.
\begin{eqnarray}
    \# \left\{x \in \{0,1\}^n : \sum_{j \in J}x_j=r\right\}=2^{n-m}\binom{m}{r}
\end{eqnarray}
Define $\omega^{(m)}=(\omega_J(x))_{x \in \{0,1\}^n}^{\downarrow}$, which depends only on $m=\vert J\vert$ and can be written as
\begin{equation}\label{eq:definition_omega_m}
    \omega^{(m)}=\Big( \underbrace{m,\hdots, m}_{2^{n-m}\binom{m}{0}}, \underbrace{m-1,\hdots,m-1}_{2^{n-m}\binom{m}{1}}, \hdots, \underbrace{0,\hdots 0}_{2^{n-m}\binom{m}{m}}  \Big)
\end{equation}
If we apply the rearrangement inequality and Schur's Theorem, we obtain
\begin{align}
    \sum_x \omega_J(x)\hat{c}_x & \leq \sum_{i=1}^N \omega_i^{(m)}c_i^{\downarrow}\\
    &= \omega_N^{(m)}\sum_{r=1}^N c_i^{\downarrow}+\sum_{k=1}^{N-1}(\omega_k^{(m)}-\omega_{k+1}^{(m)})\sum_{i=1}^k c_i^{\downarrow}\\
    &\leq \omega_N^{(m)}\sum_{i=1}^N \lambda_i+\sum_{k=1}^{N-1}(\omega_k^{(m)}-\omega_{k+1}^{(m)})\sum_{i=1}^{k}\lambda_i\\
    &=\sum_{i=1}^N \omega_i^{(m)}\lambda_i
\end{align}
Thus,
\begin{equation}
    \sum_{j=1}^m (\lambda_j^1)^{\downarrow}\leq \sum_{i=1}^N \omega_i^{(m)}\lambda_i:= B_m(\lambda)\, .
\end{equation}

Now we consider three cases:
\begin{itemize}
    \item[1.] If $1\leq m \leq n-2$, then $2^{n-m}\geq 2^2=4$, which implies
    \begin{equation}\label{eq:B1}
        B_m(\lambda)=m(\lambda_1+\lambda_2+\lambda_3)\, ,
    \end{equation}
    since $\lambda_4=\hdots=\lambda_{N-3}=0$ and the smallest $4$ elements of $\omega^{(m)}$ are zero.

    \item[2.] If $m=n-1$, then $2^{n-m}=2$, so by the form of $\omega^{(m)}$ in \eqref{eq:definition_omega_m} we obtain that 
    \begin{equation}\label{eq:B2}
        B_{n-1}(\lambda)=(n-1)\lambda_1+(n-1)\lambda_2+(n-2)\lambda_3+\lambda_{N-2}\, .
    \end{equation}
    
    \item[3.] If $m=n$, then $2^{n-m}=1$ and again looking at the form of  $\omega^{(m)}$ in \eqref{eq:definition_omega_m} we compute
    \begin{equation}\label{eq:B3}
        B_n(\lambda)=n\lambda_1+(n-1)\lambda_2+(n-1)\lambda_3+\lambda_{N-2}+\lambda_{N-1}
    \end{equation}
\end{itemize}

We compare now these values with the eigenvalues of the diagonal matrices in decreasing order. For this purpose, let $P:= \lambda_1+\lambda_2+\lambda_3$ and  $Q:= \lambda_{N-2}+\lambda_{N-1}+\lambda_N$, 
so for $1\leq j\leq n-2$,  $\tr_{[n]\setminus \{j\}}[\Lambda]$ has eigenvalues $P$ and $Q$. In addition,
\begin{equation*}
    \tr_{[n]\setminus \{n-1\}}[\Lambda]=\begin{pmatrix}
        \lambda_1+\lambda_2+\lambda_{N-2}& 0\\
        0 & \lambda_3+\lambda_{N-1}+\lambda_N
    \end{pmatrix}\, ,
\end{equation*}  
and
\begin{equation*}    
    \tr_{[n]\setminus \{n\}}[\Lambda]=\begin{pmatrix}
        \lambda_1+\lambda_3+\lambda_{N-1}& 0\\
        0 & \lambda_2+\lambda_{N-2}+\lambda_N
    \end{pmatrix}\, .
\end{equation*}
In particular, if we let $R:=\lambda_1+\lambda_2+\lambda_{N-2}$ and $S:=\lambda_1+\lambda_3+\lambda_{N-1}$, then the largest $n$ elements of the vector in the right hand-side of \eqref{eq:theo-n-partite} are $P$ a total of $n-2$ times, $R$ and $S$ with $P\geq R \geq S \geq \tr[C]/2$. As a consequence,
\begin{itemize}
    \item[1.] If $1\leq m \leq n-2$, by \eqref{eq:B1}
    \begin{equation*}
        \sum_{i=1}^{m}(\lambda_i^1)^{\downarrow} \leq mP=\sum_{j=1}^m  \lambda_i^{\downarrow}\left(\bigoplus_{j=1}^n \tr_{[n]\setminus \{j\}}[\Lambda]  \right)
    \end{equation*}
    \item [2.] If $m=n-1$, making use of \eqref{eq:B2} we obtain
    \begin{equation*}
        \sum_{i=1}^{n-1}(\lambda_i^1)^{\downarrow} \leq (n-2)P+R=\sum_{j=1}^{n-1}\lambda_i^{\downarrow}\left(\bigoplus_{j=1}^n \tr_{[n]\setminus \{j\}}[\Lambda]  \right)
    \end{equation*}
    \item[3.] If $m=n$, \eqref{eq:B3} shows that 
    \begin{equation*}
        \sum_{i=1}^{n}(\lambda_i^1)^{\downarrow} \leq (n-2)P+R+S=\sum_{j=1}^{n}\lambda_i^{\downarrow}\left(\bigoplus_{j=1}^n \tr_{[n]\setminus \{j\}}[\Lambda]  \right)
    \end{equation*}
\end{itemize}
This shows \eqref{eq:theo-n-partite2} for $1\leq m \leq n$.

By trace preservation of the partial trace, in order to conclude the proof, it is sufficient to show that
\begin{equation}
     \sum_{\substack{ j\in J \subseteq [n]\\ \vert J \vert=m } }\lambda_j^{(2)}\geq \sum_{j=1}^m  \Lambda_{[n]\setminus \{j\}}^{(2)}\, ,
\end{equation}
which follows using the first part of the proof
\begin{equation}
    \sum_{\substack{ j\in J \subseteq [n]\\ \vert J \vert=m } }\lambda_j^{(2)}=m\tr[C]-\sum_{\substack{ j\in J \subseteq [n]\\ \vert J \vert=m } }\lambda_j^{(1)}\geq m \tr[C]-\sum_{j=1}^m  \Lambda_{[n]\setminus \{j\}}^{(1)}=\sum_{j=1}^m  \Lambda_{[n]\setminus \{j\}}^{(2)}\, .
\end{equation}

\end{proof}

The next results follow after the application of Schur-convex or Schur-concave functionals, and they provide sharp bounds for all these quantities.

\begin{corollary}
    Let $C \in L((\mathbb{C}^2)^{\otimes n})$ be self-adjoint with diagonal matrix $\Lambda$ and eigenvalues ordered in the decreasing order and satisfying $\lambda_4=\hdots = \lambda_{2^n-3}$. Then for every $p\geq 1$,
    \begin{equation}
        \sum_{j=1}^n \Vert \tr_{[n] \setminus \{j\}} [C]\Vert_p^p\leq \sum_{j=1}^n \Vert \tr_{[n] \setminus \{j\} }[\Lambda] \Vert_p^p \, .
    \end{equation}
    If $0<p<1$ and $C\geq 0$ then
 \begin{equation}
        \sum_{j=1}^n \Vert \tr_{[n] \setminus \{j\}} [C]\Vert_p^p\geq \sum_{j=1}^n \Vert \tr_{[n] \setminus \{j\} }[\Lambda] \Vert_p^p \, .
    \end{equation}
\end{corollary}

\begin{corollary}
    Let $C \in L((\mathbb{C}^2)^{\otimes n})$ be a positive semidefinite matrix with diagonal matrix $\Lambda$ and eigenvalues ordered in the decreasing order satisfying $\lambda_4=\hdots = \lambda_{2^n-3}$. Then,
    \begin{equation}
        \prod_{j=1}^n\det(\tr_{[n] \setminus \{j\}} [C])\geq \prod_{j=1}^n\det(\tr_{[n] \setminus \{j\}} [\Lambda])
    \end{equation}
\end{corollary}

\begin{corollary}
    Let $\rho \in L((\mathbb{C}^2)^{\otimes n})$ be a quantum state with diagonal matrix $\Lambda$ and eigenvalues ordered in the decreasing order satisfying $\lambda_4=\hdots = \lambda_{2^n-3}$. Then,
    \begin{equation}
        \sum_{j=1}^n S( \tr_{[n] \setminus \{j\}} [\rho])\geq  \sum_{j=1}^n S( \tr_{[n] \setminus \{j\}} [\Lambda])\, ,
    \end{equation}
    where $S$ denotes the von Neumann entropy. 
\end{corollary}

\section{Conclusion}

In this paper, we have presented a unified framework to obtain sharp   bounds for Schur-convex functionals of partial traces over unitary orbits. The answer is achieved by finding majorization relations on the spectrum of partial traces, which allow the subsequent application of any  Schur-convex functional. For a single partial trace, we have shown that this approach  completely resolves the problem. Namely, for any  self-adjoint matrix, the optimal solution is given by evaluating at a diagonal matrix on the unitary orbit (in the decreasing order for $\tr_2$ and the flipped decreasing order for $\tr_1$). We then extend the reasoning to general matrices and singular values.

For a direct sum of partial traces, the situation is more complex, since we cannot always guarantee that the optimal bound can be obtained at a diagonal matrix. However, we are able to provide sufficient conditions for sharp bounds, and in addition, whenever these conditions are not satisfied, we have presented quadratic programs to obtain new bounds. To sum up,  we have presented  a framework to obtain bounds of partial traces  with the following procedure:

\begin{enumerate}
    \item Within the unitary orbit choose a covering set. In our  work we chose the set of diagonal matrices.
    \item Derive sufficient conditions on spectra to satisfy the desired majorizations.
    \item Use the sufficient conditions to produce new bounds for arbitrary spectra, as it was shown with QPs in  \cref{sec:QP}.
    \item Extend the spectral sufficient conditions to singular values of arbitrary matrices as it was shown in Section \ref{sec:SVD}.
\end{enumerate}

 There is a balance between the complexity of the covering set and the richness of the sufficient conditions on spectra. We believe diagonal matrices strike a good balance. Our framework, however, allows incremental steps in building up the covering set, since increasing the covering set can only add more sufficient conditions.  The results that our framework is based on do not depend on the specific choice of the covering set. These are Lemma \ref{lem:C_hat}, Proposition \ref{prop:diagonalization}, Remark \ref{rem:MajorizationPartialTracesByDiagLambda}, Proposition \ref{prop:MajorizationM} and Lemma \ref{lem:SVD}.

\vspace{25pt}

 \emph{\underline{Acknowledgments:}} The authors would like to thank Paul Gondolf, Michael M. Wolf and Alexander Guterman for interesting discussions. PCR acknowledges funding by the Deutsche Forschungsgemeinschaft (DFG, German Research Foundation) under Germany's Excellence Strategy –  EXC-2111 – 390814868.

\vspace{15pt}

\emph{\underline{Conflict of interest:}} The authors have no conflict of interest related to this publication.

\vspace{15pt}

\emph{\underline{Data availability:}} Data sharing is not applicable to this article as no datasets were generated or analyzed
during the current study.

\vspace{15pt}

\emph{\underline{Code availability:}} The code implementing the quadratic programs developed in Section \ref{sec:QP} and reproducing the computations in Example \ref{example:numerics} and Figures \ref{fig:pNorms}, \ref{fig:AlphaRenyi} is publicly available on \href{https://github.com/pashteiner/partial-trace-inequalities}{GitHub}. Version 1.0.0, corresponding to the computations reported in this article, is permanently archived at Zenodo \cite{QPs}.

\bibliographystyle{abbrv}
\bibliography{bibliography}

\vspace{10pt}

\appendix
\section{Proof of the sufficient condition for small dimensions}
\begin{proposition}\label{propAPP:sufficient_conditions_small_dimensions}
    Let $3 \leq d \leq 5$ and $0 \leq p, q \leq 3$ with $p+q=d$. Then, for any  self-adjoint $C \in  \L{d, d}$ with spectrum satisfying $\lambda_{p+1}=\hdots=\lambda_{d^2-q}$,
\begin{equation}\label{app:eq:d<=5:NewjointMajorizationPartialTraces}
        \lambda(\tr_1 [C] \oplus \tr_2 [C])\preceq  \lambda(\tr_1 [\Lambda] \oplus \tr_2[\Lambda]) \, .
    \end{equation}
Additionally, Condition \eqref{app:eq:d<=5:NewjointMajorizationPartialTraces} always holds for $d=2$. In other words, the following spectral conditions are sufficient for \eqref{app:eq:d<=5:NewjointMajorizationPartialTraces}:
\begin{enumerate}
    \item $d = 2$;
    \item $d = 3$ and $\lambda_n=\hdots=\lambda_{n+5}$ for some $n \in \{1,2,3,4\}$,
    \item $d = 4$ and $\lambda_{n}=\hdots=\lambda_{n+11}$ for some $n \in \{2,3,4\}$;
    \item $d = 5$ and $\lambda_{n}=\hdots=\lambda_{n+19}$ for some $n \in \{3,4\}$;
\end{enumerate}
\end{proposition}
\begin{proof}
Assume first that $d \geq 3$. The proof essentially repeats the proof of \cref{prop:NewSufficientConditionsSquareCase} with minor computational differences. Once again, we assume without loss of generality that $\lambda_{p+1}=\hdots=\lambda_{d^2-q}=0$. We can therefore write the spectrum in the form
\[\lambda(C)^T = (\lambda_1, \ldots, \lambda_p, 0, \ldots, 0, \nu_1, \ldots, \nu_q),\]
where $\lambda_1 \geq \ldots \geq \lambda_p \geq 0 \geq \nu_1 \geq \ldots \geq \nu_q$. As usual, if $p=0$ or $q=0$, the corresponding part of the vector is absent. Moreover, we may assume without loss of generality that $p \geq q$. Otherwise, we consider $-C$ and the corresponding spectrum. In particular, $p \geq 2$. Denote \[P \coloneqq \lambda_1 + \ldots + \lambda_p, \quad Q \coloneqq \nu_1 + \ldots + \nu_q.
\]
Then
    \begin{equation*}
    \lambdasum{\Lambda}^T=\Bigg(P, \lambda_1, \ldots, \lambda_p, \underbrace{0}_{d-2},\nu_1, \ldots, \nu_q, Q\Bigg), \,
\end{equation*} which is rearranged in the decreasing order.
It remains to verify that for any $k \leq 2d - 1$ condition \eqref{eq:max_k1_k2} holds for $D = \Lambda$. Note that all relevant expressions for $\sum\limits_{i=1}^{dk - \lfloor\frac{k^2}{4}\rfloor} \lambda_i + \sum\limits_{j=1}^{\lfloor\frac{k^2}{4}\rfloor} \lambda_j$ are simplified in the last column of Table~\ref{table:newSufficientConditions}.
\begin{enumerate}
    \item[$k=1$:] $\lambda^\downarrow_1(\trsum{\Lambda})=P=\sum\limits_{i=1}^{d} \lambda_i$;
    \item[$k=2$:] $\sum\limits_{i=1}^{2} \lambda^\downarrow_i(\trsum{\Lambda}) = P + \lambda_1=\sum\limits_{i=1}^{2d - 1} \lambda_i + \lambda_1$;
    \item[$k=3$:] $\sum\limits_{i=1}^{3} \lambda^\downarrow_i(\trsum{\Lambda}) = P +\lambda_1 + \lambda_2 =\sum\limits_{i=1}^{3d - 2} \lambda_i + \sum\limits_{i=1}^{2}\lambda_j$;
    \item[$4 \leq k \leq 2d-3$:\hspace{-4em}] \hspace{4em} Note that in this case $d \geq 4$.\\ Then $\sum\limits_{i=1}^{k} \lambda^\downarrow_i(\trsum{\Lambda}) = 2P = \sum\limits_{i=1}^{dk - \lfloor\frac{k^2}{4}\rfloor} \lambda_i + \sum\limits_{j=1}^{\lfloor\frac{k^2}{4}\rfloor} \lambda_j$, because
    \[2P = \sum\limits_{i=1}^{4d - 4} \lambda_i + \sum\limits_{i=1}^{4}\lambda_j \leq \sum\limits_{i=1}^{dk - \lfloor\frac{k^2}{4}\rfloor} \lambda_i + \sum\limits_{j=1}^{\lfloor\frac{k^2}{4}\rfloor} \lambda_j \leq \sum\limits_{i=1}^{d^2 - 2} \lambda_i + \sum\limits_{i=1}^{d^2 - 3d + 2}\lambda_j = 2P;\]
    \item[$k=2d-2$:\hspace{-2em}] \hspace{2em}$\sum\limits_{i=1}^{2d-2} \lambda^\downarrow_i(\trsum{\Lambda})=2P + Q - \lambda_{d^2}= (\tr[C] - \lambda_{d^2}) + P =\sum\limits_{i=1}^{d^2 - 1} \lambda_i + \sum\limits_{i=1}^{(d - 1)^2}\lambda_j$;
    \item[$k=2d-1$:\hspace{-2em}] \hspace{2em}$\sum\limits_{i=1}^{2d-1}  \lambda^\downarrow_i(\trsum{\Lambda})=2P + Q = \tr[C] + P = \sum\limits_{i=1}^{d^2} \lambda_i + \sum\limits_{i=1}^{d^2 - d}\lambda_j$.
\end{enumerate}

\medskip

Finally, in case $d = 2$ for arbitrary \[\lambda(C)^T = (\lambda_1, \lambda_2, \lambda_3, \lambda_4)\]
with $\lambda_1 \geq \lambda_2 \geq \lambda_3 \geq \lambda_4$, we obtain
    \begin{equation*}
    \lambdasum{\Lambda}^T=\Bigg(\lambda_1 + \lambda_2, \lambda_1 + \lambda_3, \lambda_2 + \lambda_4, \lambda_3 + \lambda_4\Bigg), \,
\end{equation*} which is rearranged in the decreasing order. Then it is easy to verify that for $k \leq 3$ condition \eqref{eq:max_k1_k2} holds for $D = \Lambda$:
\begin{enumerate}
    \item[$k=1$:] $\lambda^\downarrow_1(\trsum{\Lambda})=\lambda_1 + \lambda_2 =\sum\limits_{i=1}^{2} \lambda_i$;
    \item[$k=2$:] $\sum\limits_{i=1}^{2} \lambda^\downarrow_i(\trsum{\Lambda}) = 2\lambda_1 + \lambda_2 + \lambda_3 =\sum\limits_{i=1}^{3} \lambda_i + \lambda_1$;
    \item[$k=3$:] $\sum\limits_{i=1}^{3} \lambda^\downarrow_i(\trsum{\Lambda}) = 2\lambda_1 + 2\lambda_2 + \lambda_3 + \lambda_4=\sum\limits_{i=1}^{4} \lambda_i + \sum\limits_{i=1}^{2}\lambda_j$.
\end{enumerate}
\end{proof}

\section{Counterexamples for majorization by diagonal matrices}
\begin{lemma}\label{lemAPP:rankBetweenMultipliesd_2} Let $2\leq d_1 \leq d_2$ and let $C\in \L{d_1, d_2}$ be positive semidefinite. If $(k-1)d_2 < \rk(C)\leq k(d_2 - 1)$ for some $2 \leq k \leq d_1$, then there exists a unitary $U \in \U{d_1, d_2}$ such that \[\lambdatrsum{[UCU^*]}\npreceq \lambdatrsum{[\Lambda_{\pi}]} \text{ for every } \pi \in S_{d_1d_2}.\]
\end{lemma}
\begin{proof}
Let $2 \leq k \leq d_1$, $r = \rk{C} >d_2$ by assumption and consider the diagonal matrix $\Lambda_{\sigma}$ with 

\begin{equation*}
    \diag(\Lambda_{\sigma})^T=\bigoplus_{i=1}^{k-1} \underbrace{\left(\lambda_{(i-1)(d_2-1)+1},\hdots, \lambda_{i(d_2-1)},0 \right)}_{d_2} \bigoplus ( \underbrace{\lambda_{(k-1)(d_2-1) + 1},\hdots, \lambda_r, \overbrace{0, \hdots, 0}^{\geq 1}}_{d_2})
\bigoplus(\underbrace{0,\hdots,0}_{(d_1-k)d_2})
\end{equation*}

Apply now a transformation of the form \eqref{eq_T_alpha} between the diagonal entries $d_2$
 and $d_2+1$,  
\begin{equation*}
    C_{\alpha}=T_{\alpha,d_2,d_2+1}(\Lambda_{\sigma}),
\end{equation*}
where $0 < \alpha < \lambda_{r}$. Thus
\begin{align*}
\diag(C_\alpha)^T=(\underbrace{\lambda_1,\hdots, \lambda_{d_2-1},\alpha}_{d_2})&\bigoplus (\underbrace{\lambda_{d_2} - \alpha, \hdots, \lambda_{2d_2-2},0}_{d_2})\bigoplus \hdots \bigoplus\\ &\bigoplus(\underbrace{\lambda_{(k-2)(d_2-1)+1}, \hdots, \lambda_{(k-1)(d_2-1)},0}_{d_2})\\ &\bigoplus ( \underbrace{\lambda_{(k-1)(d_2-1) + 1},\hdots, \lambda_r, \overbrace{0, \hdots, 0}^{\geq 1}}_{d_2})
\bigoplus(\underbrace{0,\hdots,0}_{(d_1-k)d_2}).
\end{align*}
Note that $\lambda(C_\alpha) = \lambda(C)$ and $\tr_1 [C_{\alpha}]$, $\tr_2 [C_{\alpha}]$ are diagonal by \eqref{eq:partial_traces_T_alpha_ diagonal}.
 
 If $\lambdasum{C_{\alpha}}\preceq \lambdasum{\Lambda_{\pi}}$ for some permutation $\pi \in S_{d_1d_2}$, then $\max(\lambdasum{\Lambda_{\pi}}) > \lambda_1+\hdots + \lambda_{d_2-1}$, i.e. $\lambdasum{\Lambda_\pi}$ must contain an entry which is a sum of $d_2$ nonzero eigenvalues of $C$. Without loss of generality we may assume that this entry is a part of $\lambda^{(2)}{(\Lambda_{\pi})}$. Otherwise $d_1 = d_2$ and we can consider $F^*\Lambda_\pi F$ instead of $\Lambda_\pi$, see \eqref{eq:RelationPartialTracesFlip}. It follows that in $\Lambda_\pi$ there must be  a diagonal block of nonzero diagonal entries (see Equation \eqref{eq:computationPartialTrace}), and consequently, every element of $\lambdaone(\Lambda_{\pi})$ must be strictly positive. Since $r>(k-1)d_2$, the number of nonzero entries in $\lambdatwo(\Lambda_{\pi})$ is at least $\lceil\frac{r}{d_2}\rceil > k - 1$. Thus  $\lambdasum{\Lambda_{\pi}}$ contains at most $d_1-k$ zeros. This number has to be exact, since this is the number of zeros in $\lambdasum{C_{\alpha}}$ and $\lambdasum{C_{\alpha}}\preceq \lambdasum{\Lambda_{\pi}}$.
 
 After discarding these zeros according to \cref{rem:a<b=>ac<bc}, and observing that $\alpha$ is an entry of $\lambdasum{C_{\alpha}}$, majorization implies
 \begin{equation}
      \alpha\geq \min_{i}\left\{ \lambda_i^{(1,2)}(\Lambda_{\pi}): \lambda_i^{(1,2)}(\Lambda_{\pi})\neq 0\right\} \geq \lambda_r,
 \end{equation}
which contradicts the choice of $\alpha$.
\end{proof}

\begin{lemma}[Low rank majorization]\label{lemaAPP:r<=d_2}
    Let $2\leq d_1 < d_2$,  and let $C \in \L{d_1, d_2}$ be  a positive semidefinite  matrix with $4\leq \rk(C)\leq d_2$. Then there exists a unitary $U\in \U{d_1, d_2}$ such that \begin{equation*}\lambdatrsum{[U CU^*]}\npreceq \lambdatrsum{[\Lambda_{\pi}]} \text{ for every } \pi \in S_{d_1d_2}.\end{equation*}
\end{lemma}
\begin{proof}
    Let $r = \rk(C)$ and consider the diagonal matrix $\Lambda_{\sigma}$ with diagonal
    \begin{equation}
        \diag(\Lambda_{\sigma})^T=\left( \lambda_1, \hdots, \lambda_{r-2}, \underbrace{0, \hdots,0}_{d_2-r+2}\right)\bigoplus \left( \lambda_{r-1}, \lambda_{r}, \underbrace{0, \hdots,0}_{d_2-2}\right) \bigoplus \left( \bigoplus_{i=1}^{d_1-2}(\underbrace{0,\hdots ,0}_{d_2}) \right)\, .
    \end{equation}
    Let $C_{\alpha}$ be the matrix that results after applying a transformation of the form \eqref{eq_T_alpha}
        \begin{equation*}
        C_{\alpha}= T_{\alpha, d_2,d_2+2}\left( \Lambda_{\sigma} \right)\, ,
    \end{equation*}
where $0 < \alpha < \lambda_{r}$. 
Thus 
\begin{equation*}
\diag(C_\alpha)^T=\left( \lambda_1, \hdots, \lambda_{r-2}, \underbrace{0, \hdots,0}_{d_2-r+1},\alpha\right)\bigoplus \left( \lambda_{r-1}, \lambda_{r}-\alpha, \underbrace{0, \hdots,0}_{d_2-2}\right) \bigoplus \left( \bigoplus_{i=1}^{d_1-2}(\underbrace{0,\hdots ,0}_{d_2}) \right).
    \end{equation*}

Recall that $\lambda(C_\alpha) = \lambda(C)$ and $\tr_1 [C_{\alpha}]$, $\tr_2[C_{\alpha}]$ are diagonal by \eqref{eq:partial_traces_T_alpha_ diagonal}.
Since $r>3$, $\lambda^{(1)}(C_{\alpha})$ has $d_2 - r + 1$ zeros, while $\lambda^{(2)}(C_\alpha)$ has $d_1 - 2$ zeros. Therefore the number of zero entries of $\lambdasum{C_\alpha}$ is $d_2 + d_1 -r - 1$. 

Assume that there exists $\pi \in S_{d_1d_2}$ such that $\lambdasum{C_\alpha} \preceq \lambdasum{\Lambda_{\pi}}$. Then since $\alpha>0$, \[\lambda_1 + \ldots + \lambda_{r-2} < \lambda_1 + \ldots + \lambda_{r-2} + \alpha = \max(\lambdasum{C_{\alpha}}).\] The majorization condition imposes that $\max(\lambdasum{C_{\alpha}})\leq \max(\lambdasum{\Lambda_{\pi}})$, and consequently it follows that $\max(\lambdasum{\Lambda_{\pi}})$ must be  a sum of at least $r-1$ eigenvalues of $C$.  Depending on the position of the last ($r$-th) eigenvalue, observe that  $\lambdasum{\Lambda_{\pi}}$ is one of the vectors $a, b, c$ of \cref{lem:d2-1} padded with $d_1 + d_2 - (r + 2)$ zeros. Moreover, the latter vector is $\lambdasum{\Lambda}$.  Then by \cref{lem:d2-1}, $\lambdasum{C_{\alpha}}\preceq \lambdasum{\Lambda_{\pi}} \preceq \lambdasum{\Lambda}$. Observe that both $\lambdasum{C_{\alpha}}$ and $\lambdasum{\Lambda}$ have $d_1 + d_2 - r - 1$  zeros, which we can ignore by \cref{rem:a<b=>ac<bc}, since they do not affect the majorization. But then for the minimal non-zero  elements, the majorization implies
\begin{equation}
    \min \{ x \in \lambdasum{C_{\alpha}}: x \neq 0\}=\alpha \geq \lambda_r =\min \{ x \in \lambdasum{\Lambda}: x \neq 0\}
\end{equation}
which is a contradiction.
\end{proof}

\begin{lemma}\label{lem:d2-1}
    Let $\lambda \in \mathbb{R}^r$, $\lambda \geq 0$ and  $\sigma \in S_r$. If we let
    \begin{align*}
        a=& (\lambda_{\sigma(1)} + \ldots + \lambda_{\sigma(r-1)},
            \lambda_{\sigma(1)} , \lambda_{\sigma(2)} , \hdots , \lambda_{\sigma(r)} , \lambda_{\sigma(r)})^T\, , \\
        b=& (\lambda_{\sigma(1)} + \ldots + \lambda_{\sigma(r-1)},
            \lambda_{\sigma(1)} + \lambda_{\sigma(r)} , \lambda_{\sigma(2)} , \hdots , \lambda_{\sigma(r)} , 0)^T\, ,\\
        c=&(\lambda_1 + \ldots + \lambda_r,
            \lambda_1 , \lambda_2 , \hdots , \lambda_r , 0)^T \, .
    \end{align*}
    Then, $a \preceq b \preceq c$.
\end{lemma}
\begin{proof}
    Since $\lambda \geq 0$, it is obvious that \begin{equation*}
    (\lambda_{\sigma(1)},\lambda_{\sigma(r)})^T
       \preceq (
            \lambda_{\sigma(1)} + \lambda_{\sigma(r)}, 0
       )^T\, .\end{equation*} It follows that $a\preceq b$. Now let us prove that $b \preceq c$. If we discard repeated entries of $b$ and $c$ we obtain subvectors \begin{equation*}b'=(
            \lambda_{\sigma(1)} + \ldots + \lambda_{\sigma(r-1)},
            \lambda_{\sigma(1)} + \lambda_{\sigma(r)}
       )^T \quad \text{and}  \quad 
        c'= (
            \lambda_1 + \ldots + \lambda_r,
            \lambda_{\sigma(1)}
    )^T \, .\end{equation*} Clearly, $b' \preceq c'$ and applying  \cref{rem:a<b=>ac<bc} we conclude that $b \preceq c$.
\end{proof}

\end{document}